\documentclass[12pt]{article}

\usepackage{latexsym}
\usepackage{bbm}
\usepackage{amsopn,amscd}
\usepackage{amsfonts}
\usepackage{amssymb}
\usepackage{amsthm}
\usepackage{amsmath}
\usepackage{epsfig,graphics}
\usepackage{a4wide}
\usepackage[square,authoryear]{natbib}

\newtheorem{Theorem}{Theorem}[section]
\newtheorem{Proposition}[Theorem]{Proposition}
\newtheorem{Lemma}[Theorem]{Lemma}

\theoremstyle{remark}
\newtheorem{Remark}[Theorem]{Remark}

\def\zz{z}
\def\empha{\em}

\def\pemphas{}
\def\scsc{\sc}
\newtheorem{thm}{Theorem}[section]

\newtheorem{prop}[thm]{Proposition}
\newtheorem{defi}[thm]{Definition}

\long
\def\MSC#1\EndMSC{\def\arg{#1}\ifx\arg\empty\relax\else
      {\par\narrower\noindent%
      2000 Mathematics Subject Classification: #1\par}\fi}

\long
\def\KEY#1\EndKEY{\def\arg{#1}\ifx\arg\empty\relax\else
    {\par\narrower\noindent%
      Keywords and Phrases: #1\par}\fi}

\newcommand{\comment}[1]{}
\newcommand{\todo}[1]{}

\newcommand{\halfsum}{\vr}

\newcommand{\se}{\mr{se}}
\newcommand{\ce}{\mr{ce}}
\newcommand{\vanifac}{f}
\newcommand{\vanisum}{F}
\newcommand{\inco}{\nu}

\newcommand{\prin}{1}
\newcommand{\sing}{0}
\newcommand{\pha}{{\mc P}}
\newcommand{\cfg}{{\mc X}}

\newcommand{\Hi}{\mc{H}}
\newcommand{\group}{K}
\newcommand{\lieal}{\mf{k}}

\newcommand{\tinco}{{\tilde{\nu}}}
\newcommand{\coco}{g}
\newcommand{\vol}{\mr{vol}}
\newcommand{\scapro}[2]{\langle #1,#2 \rangle}
\newcommand{\scaproC}[2]{\langle #1,#2 \rangle}
\newcommand{\scale}{\beta}
\newcommand{\ratio}{\tau}
\newcommand{\vani}{\mc{V}}
\newcommand{\even}{\mr{g}}
\newcommand{\odd}{\mr{u}}
\newcommand{\ket}[1]{{|#1\rangle}}

\newcommand{\mc}[1]{\mathcal{#1}}
\newcommand{\mr}[1]{\mathrm{#1}}
\newcommand{\SU}{\mr{SU}}
\newcommand{\su}{\mr{su}}

\newcommand{\SL}{\mr{SL}}
\newcommand{\GL}{\mr{GL}}

\renewcommand{\Re}{{\mr{Re}}}

\newcommand{\ol}[1]{\overline{#1}}
\newcommand{\tr}{\mathrm{tr}}

\newcommand{\RR}{\mathbb{R}}
\newcommand{\CC}{\mathbb{C}}

\newcommand{\II}{\mathbbm{1}}
\newcommand{\ctg}{\mr T^\ast}
\newcommand{\tg}{\mr T}
\newcommand{\diag}{{\rm diag}}

\newcommand{\mf}{\mathfrak}

\newcommand{\U}{{\mr U}}
\newcommand{\Ad}{{\mr{Ad}}}

\newcommand{\rref}[1]{{\rm \ref{#1}}}
\newcommand{\vp}{\varphi}
\newcommand{\vr}{\varrho}
\newcommand{\ve}{\varepsilon}
\newcommand{\ble}{\begin{Lemma}}
\newcommand{\ele}{\end{Lemma}}
\newcommand{\bpp}{\begin{Proposition}}
\newcommand{\epp}{\end{Proposition}}
\newcommand{\bpf}{\begin{proof}}
\newcommand{\epf}{\end{proof}}
\newcommand{\btm}{\begin{Theorem}}
\newcommand{\etm}{\end{Theorem}}
\newcommand{\bre}{\begin{Remark}\rm}
\newcommand{\ere}{\end{Remark}}
\newcommand{\beq}{\begin{equation}}
\newcommand{\eeq}{\end{equation}}
\newcommand{\beqa}{\begin{eqnarray}}
\newcommand{\eeqa}{\end{eqnarray}}
\newcommand{\beqast}{\begin{eqnarray*}}
\newcommand{\eeqast}{\end{eqnarray*}}

\newcommand{\marke}[3]{\put(#1){\put(0.05,0.1){\makebox(-0.1,-0.2)[#2]{$#3$}}}}

\numberwithin{equation}{section}

\begin{document}

\title{\bf A gauge model for quantum mechanics\\ on a stratified space}

\author{
J.~Huebschmann$^1$, G.~Rudolph$^2$, M.~Schmidt$^2$
\\[0.3cm]
$^1$ USTL, UFR de Math\'ematiques\\
CNRS-UMR 8524
\\
59655 Villeneuve d'Ascq C\'edex, France\\
Johannes.Huebschmann@math.univ-lille1.fr
\\[0.3cm]
$^2$ Institute for Theoretical Physics, University of Leipzig\\
Augustusplatz 10/11, 04109 LEIPZIG, Germany
\\[0.3cm]
 }

\maketitle

\begin{abstract}

\noindent 
In the Hamiltonian approach on a single spatial
plaquette, we construct a quantum (lattice) gauge theory which
incorporates the classical singularities. The reduced phase space is
a stratified K\"ahler space, and we make explicit the requisite
singular holomorphic quantization procedure on this space. On the
quantum level, this procedure yields a costratified Hilbert
space, that is, a Hilbert space together with a system which
consists of the
subspaces associated with the strata of the reduced phase space and
of the corresponding orthoprojectors. The costratified Hilbert space
structure reflects the stratification of the reduced phase space.
For the special case where the structure group is
$\mathrm{SU}(2)$, we discuss the tunneling probabilities between
the strata, determine the energy eigenstates and study the
corresponding expectation values of the orthoprojectors onto the
subspaces associated with the strata in the strong and  weak
coupling approximations.

\end{abstract}

\newpage

\tableofcontents


\section{Introduction}


According to {\scsc Dirac}, the correspondence between a
classical theory and its quantum counterpart should be based on an
analogy between their mathematical structures. An interesting
issue is then that of the role of singularities in quantum
problems. Singularities are known to arise in classical phase
spaces. For example, in the Hamiltonian picture of a theory,
reduction modulo symmetries leads in general to singularities on
the classical level. Thus the question arises whether, on the
quantum level, there is a suitable structure having the classical
singularities  as its shadow and whether and how we can uncover
it. As far as we know, one of the first papers in this topic is
that of {\scsc Emmrich and R\"omer\/} \cite{emmroeme}. This paper
indicates that wave functions may \lq\lq congregate\rq\rq\ near a
singular point, which goes counter to the sometimes quoted
statement that singular points in a quantum problem are a set
of measure zero so cannot possibly be important. In a similar
vein, {\scsc Asorey et al\/} observed that vacuum nodes correspond
to the chiral gauge orbits of reducible gauge fields with
non-trivial magnetic monopole components \cite {asfalolu}. It is
also noteworthy, cf.\ e.g.\ \cite{armcusgo} and the references
there, that in classical mechanics and in classical field theories
singularities in the solution spaces are the rule rather than
the exception. This is in particular true for Yang-Mills
theories and for Einstein's gravitational theory; see for example
\cite{armamonc,armamotw}.

In \cite{kaehler}, one of us isolated a certain class
of K\"ahler spaces with singularities, referred to as stratified K\"ahler 
spaces. To explore the potential impact of classical
phase space singularities on quantum problems, in \cite{qr}, he then developed 
the notion of costratified Hilbert space. This is the
appropriate quantum state space over a stratified space; it
consists of a system of Hilbert spaces, one for each stratum which
arises from quantization on the closure of that stratum, the
stratification provides  bounded linear operators between these
Hilbert spaces reversing the partial ordering among the strata,
and these linear operators are compatible with the quantizations.
The notion of costratified Hilbert space is, perhaps, the
quantum structure which has the classical singularities as its
shadow.
In \cite{qr}, the ordinary K\"ahler quantization scheme has been
extended to such a scheme over (complex analytic) stratified
K\"ahler spaces. The appropriate quantum Hilbert space is, in
general, a costratified Hilbert space. Examples abound; one such
class of examples, involving holomorphic nilpotent orbits and in
particular angular momentum zero spaces, has been  treated in
\cite{qr}.

Gauge theory in the Hamiltonian approach, phrased on a finite
spatial lattice, leads to tractable finite-dimensional models for
which one can analyze the role of singularities explicitly. Under
such circumstances, after a choice of  tree gauge has been made,
the unreduced classical phase space amounts to the total space
$\ctg(\group\times\cdots\times\group)$ of the cotangent bundle
on a product of finitely many copies of the manifold underlying the
structure group $\group$. Gauge transformations
are then given by the lift of the action of $\group$ on
$\group\times\cdots\times\group$ by diagonal conjugation. This leads to a
finite-dimensional Hamiltonian system with symmetries. For first results on the
stratified structure of both the reduced configuration space and the reduced
phase space of systems of this type, see 
\cite{cfg,cfgtop,cdy,locpois,modus}. 
Within  canonical quantization for the unreduced system, the
algebra of observables and its representations have been
extensively investigated, see  \cite{qed1,qed2,qed3} for quantum
electrodynamics and \cite{qcd1,qcd2,qcd3} for quantum
chromodynamics. However, in this approach, the implementation of
singularities is far from being clear. 

In the present paper we will consider the case of one copy of $\group$. This
corresponds to a lattice consisting of a single plaquette. 
The unreduced phase space $\ctg \group$ carries an invariant 
complex structure, and the complex and cotangent bundle symplectic structures
combine to give an invariant K\"ahler
structure. Thus, the stratified K\"ahler quantization scheme of 
\cite{qr} referred to above can be
applied. We construct the costratified Hilbert space on the
reduced phase space by reduction after quantization. Ordinary
half-form K\"ahler quantization on $\ctg \group$ yields a Hilbert
space of holomorphic and, therefore, continuous wave
functions on $\ctg \group$, and we take the total Hilbert space of
our theory to be the subspace of $\group$-invariants. Given a
stratum, we then consider the space of functions in the Hilbert
space which vanish on the stratum, and we take the orthogonal
complement of this space as the Hilbert space associated with the
stratum. Now, in the K\"ahler polarization, among the classical
observables, only the constants can be quantized directly.
However, the holomorphic Peter-Weyl theorem \cite{holopewe} or,
equivalently, a version of the Segal-Bargmann transform
\cite{bhallone}, yields an isomorphism between the total Hilbert
space arising from K\"ahler quantization and the Hilbert space of
the Schr\"odinger representation. Via this isomorphism, the
costratified structure passes to the Schr\"odinger picture. On the
other hand, observables defined in the Schr\"odinger picture via
half-form quantization, for example, the Hamiltonian, can be
transferred to the holomorphic picture as well. Our approach
includes the quantization of arbitrary conjugation invariant
Hamiltonian systems on the total space of the cotangent bundle of
a compact Lie group. In this paper we concentrate on the
particular case of $\SU(2)$ with a lattice gauge theoretic
Hamiltonian.

The paper is organized as follows. In Section \rref{Sclassical} we
introduce the model and give a brief description of the stratified
K\"ahler structure of its reduced classical phase space. Section
\rref{Squantum} contains the construction of the costratified
Hilbert space structure for general $\SU(n)$. In Section
\rref{SSU2}, we then make this construction explicit for $\SU(2)$.
In Section \rref{Seigen} we determine the energy eigenvalues and
eigenstates of our model for $\SU(2)$. Finally, in Section
\rref{Sexpect}, we discuss the corresponding expectation values of
the orthoprojectors onto the subspaces associated with the strata
and derive approximations for strong and weak coupling.


\section{The classical picture}
\label{Sclassical}



\subsection{The model}


Let $\group$ be a compact connected Lie group
and let $\lieal$ be its Lie algebra. We consider
lattice gauge theory with structure group $\group$ in the
Hamiltonian approach on a single spatial plaquette. By means of a
tree gauge\footnote{For an arbitray lattice $\Lambda$, 
a tree gauge amounts to a choice of maximal tree in $\Lambda$,
the parallel transporters along the on-tree links being set equal to the 
identity of $\group$; this leaves the  
parallel transporters along the off-tree links as variables and constant gauge
transformations as symmetries. In our simple example, there is only 
one off-tree
link.}, the reduced phase space of the system can be shown to
be isomorphic, as a stratified symplectic space, to the reduced
phase space of the following simpler system. The unreduced
configuration space is the group manifold $\group$ and 
gauge transformations are given by the action 
of $\group$ upon itself by inner automorphisms. 
The unreduced phase space is the
cotangent bundle $\ctg\group$, acted upon by the lifted action.
This action is well known to be Hamiltonian and the corresponding
momentum mapping $\mu\colon\ctg\group\to\lieal^\ast$ is given by a
familiar expression \cite{AbrahamMarsden}. We trivialize
$\ctg\group$ in the following fashion: Endow $\lieal$ with an
invariant positive definite inner product
$\langle\cdot,\cdot\rangle$; we could take, for example, the
negative of the Killing form, but this is not necessary. By means
of the inner product, we identify $\lieal$ with its dual
$\lieal^*$ and the total space $\tg \group$ of the tangent bundle
of $\group$  with the total space $\ctg \group$ of the cotangent
bundle of $\group$. Composing the latter identification with the
inverse of left translation we obtain a diffeomorphism
 \beq\label{Gtrviz}
\ctg\group \to \tg\group \to \group\times\lieal\,.
 \eeq
It is $\group$-bi-inivariant w.r.t.\ the action of
$\group\times\group$ on $\group\times\lieal$ given by
$$
(x,Y) \mapsto (axb,\Ad_{b^{-1}} Y)
 \,,~~~~~~
a,b,x\in \group,~~ Y\in \lieal\,.
$$
In the variables $(x,Y)\in\group\times\lieal$, the lifted action of $K$ reads
$$
(x,Y) \mapsto (axa^{-1},\Ad_a Y)
 \,,~~~~~~
a\in \group\,,
$$
and the symplectic potential $\theta \colon \tg\ctg \group\to \RR$ is given by
 \beq\label{GsplPot}
\theta_{(x,Y)}(x V,W) = \langle Y,V \rangle
 \,,~~~~~~
V,W\in\lieal\,,
 \eeq
where the association $(x,V)\mapsto x V$ ($x\in \group,\, V \in
\lieal$) refers to left translation in $\mathrm T\group$. Accordingly, the
symplectic form $\omega = \mr -d \theta$ has the explicit description 
$$
\omega_{(x,Y)} \big((xV_1,W_1),(xV_2,W_2)\big)
 =
\big\langle V_1,W_2 \big\rangle
 -
\big\langle W_1,V_2 \big\rangle
 +
\big\langle Y,[V_1,V_2] \big\rangle
 \,,
$$
where $V_1,V_2,W_1,W_2\in\lieal$. The Poisson bracket of functions 
$f,g\in C^\infty(\group\times\lieal)$ is given by
 \beq\label{GPoibra}
\{f,g\}(x,Y)
 =
\big\langle f_\group(x,Y), g_\lieal(x,Y) \big\rangle
 -
\big\langle f_\lieal(x,Y), g_\group(x,Y) \big\rangle
 -
\big\langle Y , [f_\lieal(x,Y),g_\lieal(x,Y)] \big\rangle
 \,,
 \eeq
where $f_\group$ and $f_\lieal$ are $\lieal$-valued functions on
$\group\times\lieal$ representing the partial derivatives of $f$
along $\group$ and $\lieal$, respectively. They are defined by 
$$
\big\langle f_\group(x,Y) , Z \big\rangle
 =
\left.\frac{\mr d }{\mr d t}\right|_{t=0} f(x\mr e^{t Z},Y)
 \,,~~~~~~
\big\langle f_\lieal(x,Y) , Z \big\rangle
 =
\left.\frac{\mr d }{\mr d t}\right|_{t=0} f(x,Y + tZ)\,,
$$
for any $Z\in\mf g$. The momentum mapping $\mu$ takes the form
 \beq\label{Gmomap}
\mu(x,Y)=\Ad_x Y - Y, \ x\in \group,\, Y \in \lieal\,.
 \eeq
In \cite{cfg,cfgtop,cdy}, $\ctg\group$ has been trivialized by
right translation and the sign conventions necessarily differ. The
(classical unreduced) Hamiltonian $H\colon \ctg\group \to \mathbb
R$ of our model is given by
 \beq
 \label{GHaFn}
H(x,Y)
 =
\frac{1}{2} |Y|^2
 +
\frac{\inco}{2}
 \left(3 - \Re\,\tr(x)\right), \ x\in \group,\, Y \in \lieal\,.
 \eeq
Here $|\cdot|$ denotes the norm defined by the inner product on
$\lieal$, the constant $\nu$ is defined by $\inco = 1/\coco^2$, where $\coco$ is
the coupling constant, and the trace refers to some representation; below 
we will suppose $\group$ to be realized as a closed subgroup of some unitary group
$\U(n)$. Moreover, we have set the lattice spacing equal
to $1$. The Hamiltonian $H$ is manifestly gauge invariant.

\bre \label{Remmodel}

Ordinary Yang-Mills theory on $S^1$ proceeds by reduction relative
to the group of all gauge transformations. As an intermediate
step, one can perform reduction relative to the group of based
gauge transformations. This procedure provides our unreduced
model, i.~e., the Hamiltonian $\group$-space $\ctg \group$. Thus,
this model recovers a true continuum theory. Starting at the
lattice theory on a single plaquette, we have bypassed the
reduction relative to the group of based gauge transformations.
Our model therefore includes the continuum theory on $S^1$ and
serves as a building block of a lattice gauge theory as well.

The quantization of Yang-Mills theory on $S^1$ in the Hamiltonian
approach has been worked out in
\cite{Dimock, DriverHall, Hall:Coherent, Hetrick, LandsmanWren:Theta,
LandsmanWren:Coherent, Wren:Rieffel, Wren:ThetaII}. In
\cite{LandsmanWren:Theta,LandsmanWren:Coherent,Wren:ThetaII} the
authors proceed through Rieffel induction, starting from the full
continuum theory, and arrive at the Hilbert space $L^2(\group,\mr
d x)^\group$ of square-integrable functions on $\group$ invariant
under inner automorphisms of $\group$. See also \cite[\S\S
IV.3.7,8]{Landsman} and the references there. We shall arrive at
the same Hilbert space almost immediately, as we start at a later
stage in the reduction procedure, but this is only a preliminary
stage for what we are aiming at: the construction of a costratified Hilbert
space to study the role of singularities in the quantum theory. 

\ere


\subsection{The K\"ahler structure on the unreduced phase space}
\label{SSKaestr}


We recall that a K\"ahler manifold is a complex manifold, endowed
with a positive definite Hermitian form whose imaginary part,
necessarily an ordinary real 2-form, is closed and non-degenerate
and hence a symplectic structure. Equivalently, a K\"ahler
manifold is a smooth manifold, endowed with a complex and a
symplectic structure, and the two structures are required to be
compatible. One way of phrasing the compatibility condition is to
require that Poisson brackets of holomorphic functions be zero.

The unreduced phase space $\ctg\group$ acquires a K\"ahler
structure in the following manner: 
We suppose $\group$ realized as a closed subgroup of some unitary group
$\U(n)$; then the complexification $\group^{\mathbb C}$ of $\group$
is the complex subgroup of $\GL(n,\CC)$ generated by $K$.
By restriction, the polar
decomposition map
$$
\U(n) \times \mr u(n) \longrightarrow \GL(n,\CC)
 \,,~~~~~~
(x,Y) \longmapsto x \mr e^{iY}\,,
$$
yields a diffeomorphism
\begin{equation}\label{polar}
\group\times \lieal \longrightarrow \group^{\mathbb C}
 \,,~~~~~~
(x,Y)\longmapsto x\,\mathrm e^{iY}\,,
\end{equation}
commonly referred to as the polar decomposition of
$\group^{\mathbb C}$. The polar decomposition is manifestly
$\group$-bi-invariant w.r.t.\ the action of $\group\times\group$
on $\group\times\lieal$ spelled out above. Thus, the composite of
the trivialization \eqref{Gtrviz} of $\ctg\group$ with the polar
decomposition map \eqref{polar} is a $\group$-bi-invariant
diffeomorphism $\ctg\group\to\group^\CC$. The resulting complex
structure on $\ctg\group \cong \group^\CC$ and the
cotangent bundle symplectic structure combine to give 
a $\group$-bi-invariant
K\"ahler structure, having as global K\"ahler potential the real
analytic function $\kappa$ given by
\begin{equation}
\kappa(x\,\mathrm e^{iY}) =|Y|^2. \label{kappa}
\end{equation}
An explicit calculation which justifies this assertion may be
found in \cite{bhallone}.


\subsection{Symmetry reduction}
\label{SSsymred}


Let $\cfg$ denote the adjoint quotient $\group/\Ad$; this is the
reduced configuration space of our model. In the standard manner,
we decompose $\cfg$ as a disjoint union $\cfg = \bigcup_{\tau,i}
\cfg_{\tau,i}$. Here, $\tau$ ranges over the orbit types of the
action, $\cfg_{\tau}$ denotes the subset of $\cfg$ which
consists of orbits of type $\tau$, and $i$ labels the
connected components of this subset. We will refer to this
decomposition as the orbit type stratification of $\cfg$.
It is a stratification in the sense of e.~g.\ Goresky-MacPherson
\cite{Goresky}. For our purposes it suffices to know that it is a
manifold decomposition in the ordinary sense, i.\ e., the
$\cfg_{\tau,i}$ are manifolds and the frontier condition holds,
viz. $\cfg_{\tau_1,i_1}\subseteq\ol{\cfg_{\tau_2,i_2}}$ whenever
$\cfg_{\tau_1,i_1}\cap\ol{\cfg_{\tau_2,i_2}}\neq\emptyset$. An
explicit description of $\cfg$ arises from a choice of a maximal
toral subgroup $T\subseteq\group$. Let $W$ be the Weyl group of
$\group$. It is well known that the inclusion $T\hookrightarrow
\group$ induces a homeomorphism from the orbit space $T/W$ onto
the quotient $\cfg=\group/\Ad$ which identifies orbit type strata.

The reduced phase space of our model is the zero momentum reduced
space $\mu^{-1}(0)/\group$ obtained by singular Marsden-Weinstein
reduction. We denote this space by $\pha$. It acquires a
stratified symplectic structure where, similarly to the reduced
configuration space $\cfg$, the stratification is given by the
connected components of the orbit type subsets, viz.\ $\pha =
\bigcup_{\tau,i}\pha_{\tau,i}$. An explicit description of $\pha$
is obtained as follows. Let $\mf t\subseteq\lieal$ be the Lie
algebra of $T$. Given $(x,Y)\in \group\times\lieal$, according to
\eqref{Gmomap}, the vanishing of $\mu(x,Y)$ implies that $x$ and
$Y$ commute. Hence, the pair $(x,Y)$ is conjugate to an element of
$T\times\mf t$ and the injection $T\times\mf
t\hookrightarrow\group\times\lieal$ induces a homeomorphism of
$\pha$ onto the quotient $(T\times\mf t)/W$ where $W$ acts
simultaneously on $T$ and $\mf t$. This homeomorphism identifies
orbit type strata.

In the case $\group = \SU(n)$, the torus
$T$ can be chosen as the subgroup
of diagonal matrices in $\group$. Then $\mf t$ is the subalgebra
of diagonal matrices in $\lieal$. The Weyl group $W$ is the
symmetric group $S_n$ on $n$ letters, acting on $T$ and $\mf t$ by
permutation of entries. The reduced configuration space $\cfg\cong
T/W$ amounts to an $(n-1)$-simplex and the orbit type strata
correspond to its (open) subsimplices. In particular, the orbit
types are labelled by partitions $n = n_1 + \cdots + n_k$ of $n$
where the $n_i$'s are positive integers reflecting the
multiplicities of the entries of the elements of $T$. Concerning
the reduced phase space $\pha$, the orbit types of the action of
$W$ on $T\times\mf t$ are given by partitions of $n$ again, where
the $n_i$'s now are the dimensions of the common eigenspaces
of pairs in $T\times \mf t$.

\begin{figure}

 \begin{center}

\unitlength1cm

 \begin{picture}(6,3)
 \put(1,0.27){
 \marke{0,0}{tr}{\pha_+}
 \marke{4,0}{tl}{~\pha_-}
 \marke{4,2}{tl}{~\pha_\prin}
 \put(0,0){\circle*{0.2}}
 \put(4,0){\circle*{0.2}}
 }
 \put(-1,0){
 \put(0,0){\epsfig{file=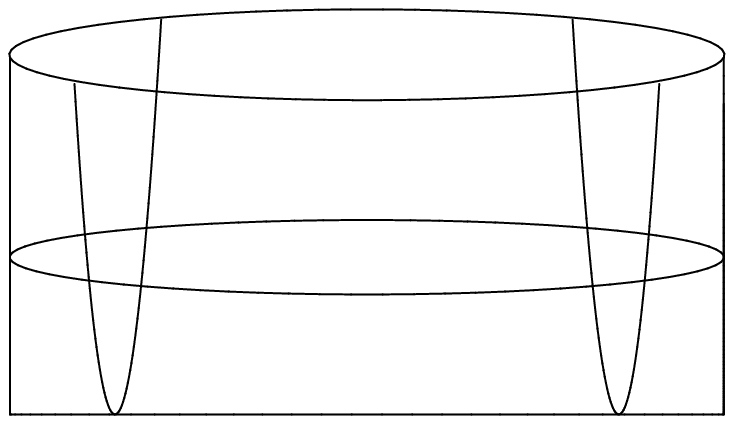,width=6cm,height=3cm}}
  }
 \end{picture}

 \end{center}

 \caption{\label{FKanu} The reduced phase space $\pha$ for
 $\group = \SU(2)$.}

 \end{figure}

For later use, we shall describe $\cfg$ and $\pha$ for $\group =
\SU(2)$ in detail. Here, $T$ amounts to the complex unit circle
and $\mf t$ to the imaginary axis. Then the Weyl group $W=S_2$
acts on $T$ by complex conjugation and on $\mf t$ by reflection.
Hence, the reduced configuration space $\cfg \cong T/W$ is
homeomorphic to a closed interval and the reduced phase space
$\pha \cong (T\times\mf t)/W$ is homeomorphic to the well-known
canoe, see Figure \rref{FKanu}. Corresponding to the partitions
$2=2$ and $2=1+1$, there are two orbit types. We denote them by
$\sing$ and $\prin$, respectively. The orbit type subset
$\cfg_\sing$ consists of the classes of $\pm\II$, i.~e., of the
endpoints of the interval; it decomposes into the connected
components $\cfg_+$, consisting of the class of $\II$, and
$\cfg_-$, consisting of the class of $-\II$. The orbit type subset
$\cfg_\prin$ is connected and consists of the remaining classes,
i.~e., of the interior of the interval. The orbit type subset
$\pha_\sing$ consists of the classes of $(\pm\II,0)$, i.~e., of
the vertices of the canoe; it decomposes into the connected
components $\pha_+$, consisting of the class of $(\II,0)$, and
$\pha_-$, consisting of the class of $(-\II,0)$. The orbit type
subset $\pha_\prin$ consists of the remaining classes, has
dimension $2$ and is connected.

\bre

In the case $\group = \SU(2)$, as a stratified symplectic space,
$\pha$ is isomorphic to the reduced phase space of a spherical
pendulum, reduced at vertical angular momentum $0$ (whence the
pendulum is constrained to move in a plane), see \cite{CuBa}.

\ere

In \cite{kaehler}, the notion of stratified K\"ahler space has
been introduced and it has been shown that, under more general
circumstances, the K\"ahler structure on $\ctg
\group\cong\group^\CC$ explained in Subsection \rref{SSKaestr}
descends to a stratified K\"ahler structure on $\pha$ which is
compatible with the stratified symplectic structure. A detailed
discussion of this stratified K\"ahler structure can be found in
\cite{adjoint,bedlewo}. For completeness, we include a brief
description in Subsection \rref{SSstrfKae} below. Since this will
not be needed for quantization, the reader who is interested in
the quantum theory only may skip this subsection.


\subsection{The stratified K\"ahler structure on the reduced phase space}
\label{SSstrfKae}


The Weyl group $W$ acts on $\ctg T$ by pull back
and on $T^\CC$ by permutation of entries.
The trivialization \eqref{Gtrviz} and the polar decomposition
\eqref{polar} combine to a $W$-equivariant diffeomorphism $\ctg T
\to T\times \mf t \to T^\CC$. This diffeomorphism,
in turn, induces a homeomorphism between $\pha$ and the quotients $\ctg
T/W\cong T^{\mathbb C}\big/W$. Moreover, as explained in
Subsection \rref{SSKaestr}, the symplectic structure of $\ctg T$
and the complex structure of $T^\CC$ combine to give a K\"ahler
structure on $\ctg T \cong T^\CC$. In the sequel, we shall stick
to the notation $T^\CC$.

Viewed as the orbit space $\ctg T\big/W$, $\pha$
inherits 
a stratified symplectic
structure by
singular
Marsden-Weinstein reduction.
That is to say: (i) The
algebra $C^\infty(T^\CC)^W$ of ordinary smooth $W$-invariant functions on
$T^\CC$ inherits a Poisson bracket and thus yields a Poisson
algebra of continuous functions on $\pha\cong T^\CC/W$, (ii) for
each stratum, the Poisson structure  yields an ordinary symplectic
Poisson structure on that stratum, and (iii) the restriction
mapping from $C^{\infty}(T^\CC)^W$ to the algebra of ordinary
smooth functions on that stratum is a Poisson map.

Viewed as the orbit space $T^\CC\big/W$,  $\pha$ acquires a
complex analytic structure in the standard fashion. 
The complex structure and the Poisson structure
combine to give a stratified K\"ahler structure on $\pha$
\cite{kaehler}, \cite{adjoint}, \cite{bedlewo}. Here the precise
meaning of the term \lq\lq stratified K\"ahler structure\rq\rq\ is
that the Poisson structure satisfies (ii) and (iii) above and that
the Poisson and complex structures satisfy the additional
compatibility requirement that for each stratum, necessarily a
complex manifold, the symplectic and complex structures on that
stratum combine to give an ordinary K\"ahler structure.

In the case $K=\SU(n)$, the complex analytic structure admits the
following elementary description: Let $\mathrm{Diag}(n,\mathbb C)$
be the group of diagonal matrices in the full linear group
$\GL(n,\CC)$. The Weyl group $W$ acts on $\mathrm{Diag}(n,\mathbb
C)$ by permutation of entries and the injection of $T^{\mathbb C}$
into $\mathrm{Diag}(n,\mathbb C)$ is compatible with this action.
The $n$ elementary symmetric functions $\sigma_1,\dots,\sigma_n$
furnish a map
\[
(\sigma_1,\dots,\sigma_n)\colon \mathrm{Diag}(n,\mathbb C)
\longrightarrow \mathbb C^n
\]
into complex $n$-space $\mathbb C^n$. The restriction
 \begin{equation}\label{proj}
(\sigma_1,\dots,\sigma_{n-1})\colon T^{\mathbb C} \longrightarrow
\mathbb C^{n-1}
 \end{equation}
 of that map  to $T^{\mathbb C}$ identifies the orbit space
$\pha\cong T^{\mathbb C}\big/W$ with the affine subspace of $\mathbb C^n$
given by the equation $\sigma_n =1$ which, in turn, may be identified with a
copy of $\mathbb C^{n-1}$. In this way, $\pha$ inherits an obvious complex
structure. Thus, affine complex $n$-space $\mathbb C^n$
appears here as the space of normalized complex degree $n$ polynomials,
and the orbit space $T^{\mathbb C}/W$ amounts to the subspace of
normalized complex degree $n$ polynomials with constant
coefficient equal to 1. Indeed, a normalized degree $n$ polynomial
$p(z)=z^n+a_{1}z^{n-1} + \ldots+ a_{n-1} z + a_n$ decomposes into
its linear factors $p(z) = \prod_j (z-z_j)$, and the coefficients
$a_j$ are given by
\[
a_{j}=(-1)^j\sigma_j(z_1,\ldots,z_n),\ 1\leq j \leq n;
\]
up to the signs  $(-1)^j$, the map $\sigma$ may thus be viewed as
that which sends the $n$-tuple $\zz_1, \ldots, \zz_n$ to the
unique normalized degree $n$ polynomial having $\zz_1, \ldots,
\zz_n$ as its zeros, the coefficients of degree $\leq n-1$ being
taken as coordinates on the space of polynomials. A more profound
analysis shows that, indeed, in terms of 
$\mathrm {SL}(n,\mathbb C)$ and $\mathrm {GL}(n,\mathbb C)$, the
passage to the quotient (which is here realized via the map
\eqref{proj}) amounts to the assignment to a matrix in $\mathrm
{SL}(n,\mathbb C)$ (or $\mathrm {GL}(n,\mathbb C)$) of its
characteristic polynomial.

We shall now describe the stratified K\"ahler structure on $\pha$
explicitly for $\group = \SU(2)$. Here, $T^\CC$ consists of the diagonal
matrices $\mathrm{diag}(z,z^{-1})$ where $z\in\CC^\ast$. The
non-trivial element of $W$ interchanges $z$ and $z^{-1}$. To determine
the complex structure we note that the map \eqref{proj} is given by
the restriction of the first elementary symmetric function
$\sigma_1$ on $\mr{Diag}(2,\CC)$ to the subgroup $T^{\mathbb C}$,
i.e.,
 \begin{equation}\label{sigmaSU2}
\sigma_1 \colon T^\CC \longrightarrow \mathbb C
 \,,~~~~~~
\sigma_1(\mathrm{diag}(z,z^{-1})) = z+ z^{-1}\,;
 \end{equation}
this map identifies $T^\CC/W\cong \pha$ with a copy of $\CC$ and thus
provides a holomorphic coordinate on $\pha$. In particular,
topologically, the canoe shown in Figure \rref{FKanu} is just an
ordinary plane.

To arrive at a description of the Poisson algebra
$C^\infty(T^\CC)^W$, we recall that, once a choice of
finitely many generators, say $p$, for the algebra
$\RR[T^\CC]^W$of real $W$-invariant polynomials on $T^\CC$ 
has been made,
the resulting Hilbert map induces 
a homeomorphism from $T^\CC/W\cong\pha$ onto a semi-algebraic
subset of $\RR^p$.
According to a theorem in \cite{Schwarz}, any
element of $C^\infty(T^\CC)^W$ can be written as a smooth function
in these generators. Hence, to describe the Poisson algebra
$C^\infty(T^\CC)^W$ it suffices to list the Poisson brackets of
these generators. In the case at hand, a set of generators for
$\RR[T^\CC]^W$ can be obtained as follows. The complexification
$\mathbb R[T^{\mathbb C}]_{\mathbb C}$ of $\mathbb R[T^{\mathbb
C}]$ is generated by $z,z^{-1},\overline z, \overline z^{-1}$.
Since the non-trivial element of $W$ interchanges $z$ and $z^{-1}$
as well as $\overline z$ and $\overline z^{-1}$, the subalgebra
$\RR[T^\CC]^W_{\mathbb C}$ of $W$-invariants is
generated by the three elementary bisymmetric functions
\[
\sigma_1 = z+z^{-1},\ \overline \sigma_1 = \overline z+ \overline
z^{-1}, \
 \sigma = z \ol z ^{-1} + \ol z z ^{-1}\,,
\]
and this algebra may be
identified with the complexification of $\RR[T^\CC]^W$ in an obvious manner.
These generators are subject to the single defining relation
 \beq\label{Grelcpx}
(\sigma_1^2-4)(\overline \sigma_1^2-4) = (\sigma_1 \overline
\sigma_1 -2\sigma)^2\,,
 \eeq
see \cite{adjoint}. Hence, $\RR[T^\CC]^W$ is generated by the
three real functions $X$, $Y$ and $\sigma$, where $\sigma_1 = X +
\mr i Y$. For convenience, instead of $\sigma$, we use $\tau =
\frac{2-\sigma}4$. In view of \eqref{Grelcpx}, the generators
$X$, $Y$, $\tau$ are subject to the relation
 \begin{equation}\label{relation}
Y^2 = (X^2 + Y^2 + 4 (\tau -1)) \tau\,.
 \end{equation}
In terms of the real coordinates $x$ and $y$ on
$T^\CC\cong\CC^\ast$ defined by $z = x + \mr i y$,
 \beq\label{GXx}
X = x + \frac{x}{r^2}
 \,,~~~~~~
Y = y - \frac{y}{r^2}
 \,,~~~~~~
\tau = \frac{y^2}{r^2}\,,
 \eeq
where $r^2 = x^2 + y^2$. The obvious inequality $\tau \geq 0$
brings the semialgebraic nature of the quotient $\mathbb
T^{\mathbb C}\big/W$ to the fore. 
To determine the Poisson brackets among the generators $X$, $Y$ and
$\tau$, we recall that, in terms of the coordinates $x$
and $y$, the symplectic structure on $T^\CC\cong \CC^\ast$ is
given by $\frac{1}{r^2} \mr d x \wedge \mr d y$ whence
\[ 
\{x,y\} = r^2.
\] 
A straightforward calculation involving
\eqref{GXx} yields the Poisson brackets
$$
\{X,Y\} = X^2 + Y ^2 + 4(2 \tau -1)
 \,,~~~~~~
\{X,\tau\} = 2(1- \tau)Y
 \,,~~~~~~
\{Y,\tau\} = 2 \tau X\,.
$$
The Poisson structure vanishes at the two points $(X,Y)=(2,0)$ and
$(X,Y)=(-2,0)$ representing the orbit type strata $\pha_+$ and
$\pha_-$, respectively. Hence, the resulting complex algebraic
stratified K\"ahler structure on $\pha$ is singular at
these two points. Furthermore, solving \eqref{relation} for
$\tau$, we obtain
\[
\tau = \frac 12 \sqrt{Y^2+ \frac {(X^2 + Y^2-4)^2}{16}} -
\frac{X^2 + Y^2 -4}{8} ,
\]
whence, at $(X,Y)=(\pm 2,0)$, $\tau$ is not smooth as a function
of the variables $X$ and $Y$. Away from these two points, i.e., on
the principal stratum $\pha_\prin$, the Poisson structure is
symplectic. We refer to the stratified K\"ahler space under
discussion as the exotic plane with two vertices. More
details and, in particular, an interpretation in terms of
discriminant varieties, may be found in \cite{bedlewo}.

\bre

The algebra $\mathbb R[T^{\mathbb C}]^W$ is the real coordinate
ring of $T^{\mathbb C}/W$, viewed as a real semi-algebraic set.
Similarly, for the description of the Poisson structure on $\pha$
we could have used a set of generators of, e.g., the algebra
$\RR[T\times\mf t]^W$ of real $W$-invariant polynomials on
$T\times\mf t$. This is the real coordinate ring of $(T\times\mf
t)/W$, viewed, in turn, as a semi-algebraic set. Since the
diffeomorphism $T\times\mf t \cong T^\CC$ is 
not algebraic, 
$\mathbb R[T^{\mathbb C}]^W$ and $\RR[T\times\mf t]^W$ correspond
to different subalgebras of the Poisson algebra
$C^\infty(T^\CC)^W$ defining the Poisson structure on $\pha\cong
T^\CC/W$.

\ere


\section{The quantum picture}
\label{Squantum}


Our aim is to push further, in the context of stratified spaces,
the ideas which underlie the program of geometric quantization. As
our physical Hilbert space we take a certain space of
square-integrable holomorphic functions which arises by K\"ahler
quantization \cite{Sniatycki,woodhous}. Through an analogue of the
Peter-Weyl theorem, this space is related with the physical
Hilbert space arising by ordinary Schr\"odinger quantization on
$\group$. Within this Hilbert space we construct the additional
structure of a costratification. Thereafter, we discuss
observables.


\subsection{Holomorphic quantization}


Let $\varepsilon$ be the symplectic (or Liouville) volume form on
$\mathrm T^*\group\cong \group^{\mathbb C}$. In terms of the polar
decomposition \eqref{polar}, we then have the identity
$\varepsilon=dxdY$ where $dx$ is the volume form on $\group$
yielding Haar measure, normalized so that it
coincides with the Riemannian volume measure on $\group$,
 and where $dY$ is the form inducing
Lebesgue measure on $\lieal$, normalized by the inner product on $\lieal$. 
Next, let $\eta$ be the real $\group$-bi-invariant analytic function on $\group
\times\lieal\cong\group^{\mathbb C}$ defined by
\[
\eta(x\,\mathrm e^{iY})
=\sqrt{\mathrm{det}\left(\frac{\sin(\mathrm{ad}(Y))}{\mathrm{ad}(Y)}\right)},
\ x \in \group, \,Y \in \lieal,
\]
the square root being the positive one. We note that $\eta^2$ is
the density of Haar measure on $\group^{\mathbb C}$ relative to
Liouville measure $\varepsilon$, cf.\
 \cite{bhallfou} (Lemma 5). To express $\eta$ in terms of a root system, we
choose a dominant Weyl chamber in the Cartan subalgebra $\mf t$ of $\lieal$
and denote by $R^+$ the corresponding set of positive roots. Then, on
$T\times\mf t\cong T^\CC$, $\eta$ is given by
$$
\eta(x\,\mathrm e^{iY})
 =
\prod\nolimits_{\alpha\in R^+} \frac{\sinh(\alpha(Y))}{\alpha(Y)}
 \,,~~~~~~
x \in T, \,Y \in \mathfrak t\,,
$$
cf.\ \cite{bhallone} (2.10). Here the $\alpha$'s are the real
roots, given by $-\mr i$ times the ordinary complex roots. Let
$\kappa$ be the $\group$-bi-invariant real analytic function on
$\group \times \lieal \cong \group^{\mathbb C}$ defined by
\eqref{kappa}. Half-form K\"ahler quantization on $\group^{\mathbb
C}$ yields the Hilbert space $\mathcal HL^2(\group^{\mathbb
C},\mathrm e^{-\kappa/\hbar}\eta \varepsilon)$ of holomorphic
functions on $\group^{\mathbb C}$ which are square-integrable
relative to the measure $\mathrm e^{-\kappa/\hbar}\eta
\varepsilon$ \cite{bhallone}. The scalar product is given by
 \beq\label{GscaproKC}
\scaproC{\psi_1}{\psi_2}
 =
 \frac{1}{\vol(\group)} \int_{\group^\CC}
\ol{\psi_1}\psi_2 \mathrm e^{-\kappa/\hbar}\eta \varepsilon\,.
 \eeq
For our purpose there is no need to write down the relevant
half-forms explicitly. They are subsumed under the measure.

Left and right translation turn the Hilbert space $\mathcal
HL^2(\group^{\mathbb C},\mathrm e^{-\kappa/\hbar}\eta
\varepsilon)$ into a unitary representation of $\group\times
\group$. The Hilbert space associated with $\pha$ by reduction
after quantization is the subspace $\mathcal HL^2(\group^{\mathbb
C},\mathrm e^{-\kappa/\hbar}\eta \varepsilon)^\group$ of
$\group$-invariants relative to conjugation.

We will now describe the Hilbert space $\mathcal
HL^2(\group^{\mathbb C},\mathrm e^{-\kappa/\hbar}\eta
\varepsilon)^\group$ as a Hilbert space of $W$-invariant
holomorphic functions on $T^{\mathbb C}$ that are
square-integrable relative to a measure of the kind $\mathrm
e^{-\kappa/\hbar}\gamma \varepsilon_T$ for a suitable density
function $\gamma$ on $T^{\mathbb C}$ where $\varepsilon_T$ denotes
the Liouville volume form on $T^{\mathbb C}\cong \mathrm \ctg T$.
This Hilbert space may in fact be viewed as coming from
quantization after reduction, i.~e., by quantization on $T^\CC/W$.
Here and below we do not distinguish in notation between the
function $\mathrm e^{-\kappa/\hbar}$ defined on $\group^{\mathbb
C}$ and its restriction to $T^{\mathbb C}$.

Let $m=\dim \group$ and $r=\dim T$. To construct the function
$\gamma$, consider the conjugation mapping
\begin{equation} q^{\mathbb C}\colon
\left(\group^{\mathbb C}\big/ T^{\mathbb C}\right) \times
T^{\mathbb C} \longrightarrow \group^{\mathbb C},\ (y T^{\mathbb
C},t)\mapsto yty^{-1},\  y\in \group^{\mathbb C}, t\in T^{\mathbb
C}, \label{conjugation}
\end{equation}
and integrate the induced $(2m)$-form $(q^{\mathbb C})^*(\mathrm
e^{-\kappa/\hbar} \eta \varepsilon)$ over \lq\lq the fibers\rq\rq\
$\group^{\mathbb C}\big/ T^{\mathbb C}$. Although the fibers are
non-compact, in view of the Gaussian constituent $\mathrm
e^{-\kappa/\hbar}$, this integration is a well defined operation.
Let $\widetilde \gamma$ be the density of the resulting
$(2r)$-form on $T^{\mathbb C}$ relative to the Liouville volume
form $\varepsilon_{T}$ on $T^{\mathbb C}\cong \mathrm T^*T$, and
let
 \begin{equation*}
\gamma= \frac{\widetilde \gamma}{|W| \mathrm e^{-\kappa/\hbar}}
 \end{equation*}
where $|W|$ is the order of the Weyl group. 
An explicit calculation of $\gamma$ can be found in Theorem 3 of Section 2 of
\cite{Florentino}, see also Theorem 12 in \cite{HallMitchell}.
The following is the
analogue of Weyl's integration formula, spelled out for $\mathrm
{Ad}(\group)$-invariant holomorphic functions.

\begin{prop}
Given two holomorphic $\mathrm {Ad}(\group)$-invariant functions
$\psi_1,\psi_2$ on $\group^{\mathbb C}$ that are square-integrable
relative to the measure $ \mathrm e^{-\kappa/\hbar} \eta \varepsilon$,
 \begin{equation}
\int_{\group^{\mathbb C}} \overline {\psi_1} \psi_2 \mathrm
e^{-\kappa/\hbar}\eta\varepsilon\,
 =
\int_{T^{\mathbb C}} \overline {\psi_1} \psi_2 \mathrm
e^{-\kappa/\hbar}\gamma\varepsilon_{T}\,.
 \end{equation}

\end{prop}

\begin{proof}
Since $\psi_1$ and $\psi_2$ are $\Ad(\group)$-invariant and holomorphic, they 
are $\Ad(\group^\CC)$-invariant. Hence, their pullbacks under the conjugation
mapping \eqref{conjugation} are constant along the constituent 
$\group^\CC/T^\CC$.
Since the conjugation mapping has degree equal to
the order $|W|$ of the Weyl group and since the complement of the image
under the conjugation mapping has measure zero,
 \begin{equation*}
\int_{\group^{\mathbb C}} \overline {\psi_1} \psi_2 \mathrm
e^{-\kappa/\hbar} \eta \varepsilon =\frac 1{|W|}\int_{T^{\mathbb
C}} \overline {\psi_1} \psi_2 \widetilde\gamma\varepsilon_{T}=
\int_{T^{\mathbb C}} \overline {\psi_1} \psi_2 \mathrm
e^{-\kappa/\hbar}\gamma\varepsilon_{T}\,.
 \end{equation*}
\end{proof}

\noindent
The proposition implies that the restriction mapping induces an isomorphism
\begin{equation}
\mathcal HL^2(\group^{\mathbb C},\mathrm e^{-\kappa/\hbar}\eta
\varepsilon)^\group
 \longrightarrow
\mathcal H L^2(T^{\mathbb
C},\mathrm e^{-\kappa/\hbar} \gamma \varepsilon_{T})^W
\end{equation}
of Hilbert spaces where, according to \eqref{GscaproKC}, the scalar
product in $\mc H L^2(T^{\mathbb C},\mathrm e^{-\kappa/\hbar}
\gamma \varepsilon_{T})^W$ is given by
 \beq\label{GscaproTC}
 \frac{1}{\vol(\group)}
 \int_{T^\CC} \ol{\psi_1}\psi_2 e^{-\kappa/\hbar}\gamma\varepsilon_{T}\,.
 \eeq

A basis of $\mathcal HL^2(\group^{\mathbb C},\mathrm
e^{-\kappa/\hbar}\eta \varepsilon)^\group$ and hence of $\mathcal
H L^2(T^{\mathbb C},\mathrm e^{-\kappa/\hbar} \gamma
\varepsilon_{T})^W$ is obtained as follows. For a highest weight
$\lambda$ relative to the chosen dominant Weyl chamber, we will
denote by $\chi^{\mathbb C}_{\lambda}$ the irreducible character
of $\group^{\mathbb C}$ associated with $\lambda$. The holomorphic
Peter-Weyl theorem established in \cite{holopewe}, 
see Remark \rref{R-SBtrf} below for historical comments, implies that the
total Hilbert space $\Hi$ contains the complex vector space which
underlies the algebra $\mathbb C[\group^{\mathbb C}]^\group$ of
$\mathrm{Ad}(\group)$-invariant polynomial functions on
$\group^{\mathbb C}$ as a dense subspace. Hence the irreducible
characters $\chi^{\mathbb C}_{\lambda}$ of $\group^{\mathbb C}$
form a basis of $\mathcal HL^2(\group^{\mathbb C},\mathrm
e^{-\kappa/\hbar}\eta \varepsilon)^\group$.


\subsection{Schr\"odinger quantization}


Half-form Schr\"odinger quantization on $\ctg\group$ yields the
Hilbert space $L^2(\group,\mr d x)$ of ordinary square-integrable
functions on $\group$ \cite{bhallone} with scalar product
 \beq\label{GscaproK}
\scapro{\psi_1}{\psi_2}
 =
\frac 1 {\vol(\group)} \int_\group \ol{\psi_1}\psi_2 \, \mr d x\,.
 \eeq
We remind the reader that for reasons explained above we have
normalized the Haar 
measure on $\group$ so that it coincides with the Riemannian volume measure. 
Left and right translation turn the Hilbert space $L^2(\group,\mr d x)$ into a
unitary 
$(\group\times \group)$-representation. The Hilbert space
associated with $\pha$ by reduction after quantization is the
subspace $L^2(\group,\mr d x)^\group$ of $\group$-invariants. It
also arises as the physical Hilbert space of the observable algebra
\cite{qcd3} and by quantization via Rieffel induction
\cite{LandsmanWren:Theta,LandsmanWren:Coherent,Wren:Rieffel,Wren:ThetaII},
see also \cite[\S\S IV.3.7,8]{Landsman}.

Similarly as $\mc H L^2(\group^\CC,\mr
e^{-\kappa/\hbar}\eta\ve)^\group$, the space $L^2(\group,\mr d x)^\group$
can alternatively be viewed as a Hilbert space of $W$-invariant
functions which now live on $T$ rather than on $T^\CC$. Indeed,
let $v\colon T\to\mathbb R$ be the real function given by
$v(t)=\mathrm{vol}(\mathrm{Ad}(\group)t)/|W|$, $t\in T$, that is,
$v(t)$ is the Riemannian volume of the conjugacy class
$\mathrm{Ad}(\group)t$ in $\group$ generated by $t\in T$, divided
by the order $|W|$ of the Weyl group. Restriction of
$\mathrm{Ad}(\group)$-invariant functions from $\group$ to $T$ is
well known to induce an isomorphism
\begin{equation}
L^2(\group,dx)^\group \longrightarrow L^2(T, v\mr d t)^W
\end{equation}
of Hilbert spaces where the scalar product on $L^2(T, vdt)^W$ is
given by
 \beq\label{GscaproT}
\frac 1 {\vol(\group)} \int_T \ol{\psi_1}\psi_2 \, v \, \mr d t\,.
 \eeq
Given a highest weight $\lambda$, we will denote by
$\chi_{\lambda}$ the corresponding irreducible character of
$\group$, so that $\chi_{\lambda}$ is the restriction of
$\chi^{\mathbb C}_{\lambda}$ to $\group$. The $\chi_\lambda$'s
form an orthonormal basis of $L^2(\group,\mr d x)^\group$.

Let $\rho=1/2\sum_{\alpha\in R^+} \alpha$ denote the half sum of the
positive roots and let $C_{\lambda}$ be the constant
 \beq\label{G-Cl}
C_{\lambda}=(\hbar\pi)^{\dim(\group)/2}\mathrm
e^{\hbar|\lambda+\rho|^2},
 \eeq
where $|\lambda+\rho|$ refers to the norm of $\lambda+\rho$
relative to the inner product on $\lieal$.

\begin{thm} \label{comparison}

The assignment to $\chi_{\lambda}$ of
$C_{\lambda}^{-1/2}\chi^{\mathbb C}_{\lambda}$, as $\lambda$
ranges over the highest weights, yields a unitary isomorphism
 \begin{equation}
L^2(\group,\mr d x)^\group
 \longrightarrow
\mc HL^2(\group^\CC,\mr e^{-\kappa/\hbar}\eta\ve)^\group
 \label{H}
 \end{equation}
of Hilbert spaces.

\end{thm}

\begin{proof}

The holomorphic function 
$C_{\lambda}^{-1/2}\chi^{\mathbb C}_{\lambda}$ is the image of $\chi_\lambda$
under the Segal-Bargmann transform
 \beq\label{Hallg}
L^2(\group,\mr d x) \longrightarrow \mc HL^2(\group^\CC,\mr
e^{-\kappa/\hbar}\eta\ve)
 \eeq
which is a unitary isomorphism \cite{Hall94}.
\todo{Gilt das exakt oder nur bis auf einen Faktor?}
The assertion follows because $\chi_\lambda$ and $\chi_\lambda^\CC$ are bases in
$L^2(\group,\mr d x)^\group$ and $\mc HL^2(\group^\CC,\mr
e^{-\kappa/\hbar}\eta\ve)^\group$, respectively. Alternatively, the assertion
is a direct consequence of Theorem 5.3 in \cite{holopewe}.

\end{proof}

\bre\label{R-SBtrf}

The Segal-Bargmann transform \eqref{Hallg} and, therefore, the isomorphism 
\eqref{H}, rely on the description of the Hilbert spaces $L^2(\group,dx)$
and $\mathcal H L^2(\group^{\mathbb C},\mathrm e^{-\kappa/\hbar}
\eta \varepsilon)$ as half-form Hilbert spaces and involve the appropriate 
metaplectic correction \cite{woodhous}.

Originally, in \cite{Hall94}, see also \cite{Hall97}, the Segal-Bargmann
transform was developed via heat kernel analysis on $\group$ and $\group^\CC$.
More recently, an alternative purely geometric description of this transform 
in terms of representative functions and independent of heat kernel 
analysis has been given in Theorem 5.3 of \cite{holopewe}. This description
relies on the holomorphic Peter-Weyl theorem \cite{holopewe}.
The holomorphic Peter-Weyl theorem yields a proof of Theorem \rref{comparison}
above as well and the geometric methods in \cite{holopewe}
also recover the heat kernel analysis. On the other hand,
the holomorphic Peter-Weyl theorem can likewise be deduced from the 
Segal-Bargmann transform developed in \cite{bhallone}, combined with the
ordinary Peter-Weyl theorem.

Alternatively, we can describe the isomorphism \eqref{Hallg} as
being induced  by the corresponding BKS-pairing map 
from $L^2(\group,\mr d x)$ to $\mc HL^2(\group^\CC,\mr
e^{-\kappa/\hbar}\eta\ve)$, multiplied by a factor 
$(4\pi)^{-\dim(\group)/4}$. For
details, see \cite{bhallone} (description in terms of the
heat kernel on $K$) or Theorem 6.5 in \cite{holopewe} (description in terms of
representative functions).

\ere

Theorem \rref{comparison} entails that the complex
characters $\chi^\CC_\lambda$ satisfy the orthogonality relations
 \beq\label{GogonrelchiC}
\scaproC{\chi^\CC_\lambda}{\chi^\CC_{\lambda'}} = C_\lambda
\delta_{\lambda\,\lambda'}\,.
 \eeq
Hence, the vectors $C_\lambda^{-1/2} \chi^\CC_\lambda$, where
$\lambda$ ranges over the highest weights, form an orthonormal
basis of $\mc HL^2(\group^\CC,\mr
e^{-\kappa/\hbar}\eta\ve)^\group$.

From now on, we will take the Hilbert space of our model 
to be the Hilbert space $\Hi$ with orthonormal basis $\ket\lambda$ labelled by
the highest weights. In the holomorphic representation, $\Hi$
is then realized as $\mc HL^2(\group^\CC,\mr
e^{-\kappa/\hbar}\eta\ve)^\group$ or, equivalently, as $\mc
HL^2(T^\CC,\mr e^{-\kappa/\hbar}\gamma\ve_T)^W$ whereas, in the
Schr\"odinger representation, $\Hi$ is realized as $L^2(\group,\mr d
x)^\group$ or, equivalently, as $L^2(T,v\,\mr d t)^W$. The passage
to the respective representation is achieved by substitution for
$\ket \lambda$ of the function $C_\lambda^{-1/2} \chi^\CC_\lambda$
or $\chi_\lambda$ as appropriate.


\subsection{The costratified Hilbert space structure}
\label{SScostrfHispa}


We will now construct the additional structure of a costratification. To begin
with, we recall from \cite{kaehler} the precise definition
of a costratified Hilbert space. Let $N$ be a
stratified space. Let $\mathcal C_N$ be the category whose objects
are the strata of $N$ and whose morphisms are the inclusions $Y'
\subseteq \overline Y$ where $Y$ and $Y'$ are strata.

\begin{defi}\label{def2}

A costratified Hilbert space relative to $N$ is a
contravariant functor from $\mathcal C_N$ to the category of
Hilbert spaces, with bounded linear maps as morphisms.

\end{defi}

In more down to earth terms, a costratified Hilbert space relative
to $N$ assigns a Hilbert space $\mathcal C_Y$ to each stratum $Y$,
together with a bounded linear map $\mathcal C_{Y_2} \to \mathcal
C_{Y_1}$ for each inclusion $Y_1 \subseteq \overline {Y_2}$ such
that, whenever $Y_1 \subseteq \overline {Y_2}$ and $Y_2 \subseteq
\overline {Y_3}$, the composite of $\mathcal C_{Y_3} \to \mathcal
C_{Y_2}$ with $\mathcal C_{Y_2} \to \mathcal C_{Y_1}$ coincides
with the bounded linear map $\mathcal C_{Y_3}\to\mathcal C_{Y_1}$
associated with the inclusion $Y_1 \subseteq \overline {Y_3}$.

To construct a costratified Hilbert space relative to the reduced
phase space $\pha$, we start with the Hilbert space $\mc HL^2(\group^\CC,\mr
e^{-\kappa/\hbar}\eta\ve)^\group$ and single out subspaces
$\Hi_{\tau,i}$ associated with the strata $\pha_{\tau,i}$ as
follows. The elements of $\mc HL^2(\group^\CC,\mr
e^{-\kappa/\hbar}\eta\ve)^\group$ are ordinary functions on
$\group^\CC$, not classes of functions as in the $L^2$-case. Therefore,
being $\group$-invariant, these functions define functions
on $\pha$. Thus, we associate with each stratum $\pha_{\tau,i}$ of $\pha$ the
subspace 
\[
\vani_{\tau,i} =\{f\mc \in\mc HL^2(\group^\CC,\mr
e^{-\kappa/\hbar}\eta\ve)^\group;\ f|_{\pha_{\tau,i}}=0 \}
\]
of $\mc HL^2(\group^\CC,\mr
e^{-\kappa/\hbar}\eta\ve)^\group$ which consists of the functions that vanish
on $\pha_{\tau,i}$. We then define the Hilbert space $\Hi_{\tau,i}$ associated
with $\pha_{\tau,i}$ to be the orthogonal complement of $\vani_{\tau,i}$ in
$\mc HL^2(\group^\CC,\mr e^{-\kappa/\hbar}\eta\ve)^\group$, so that
$\mc HL^2(\group^\CC,\mr e^{-\kappa/\hbar}\eta\ve)^\group
=\vani_{\tau,i}\oplus\Hi_{\tau,i}$.

By construction, if $\pha_{\tau_1,i_1}\subseteq\ol{\pha_{\tau_2,i_2}}$ then
$\vani_{\tau_2,i_2} \subseteq \vani_{\tau_1,i_1}$ and, therefore,
$\Hi_{\tau_1,i_1}\subseteq\Hi_{\tau_2,i_2}$. Let
$\Pi_{\tau_2,i_2;\tau_1,i_1}:\Hi_{\tau_2,i_2}\to\Hi_{\tau_1,i_1}$
denote the orthogonal projection. The resulting system
$\{\Hi_{\tau,i}\}$, together with the orthogonal projections
$\Pi_{\tau_2,i_2;\tau_1,i_1}: \Hi_{\tau_2,i_2}\to
\Hi_{\tau_1,i_1}$ whenever $\pha_{\tau_1,i_1} \subseteq \overline
{\pha_{\tau_2,i_2}}$, is the costratified Hilbert space relative to
$\pha$ we are looking for. When $\tau$ is the principal orbit type, $\Hi_\tau$
plainly coincides with the total Hilbert space $\mc HL^2(\group^\CC,\mr
e^{-\kappa/\hbar}\eta\ve)^\group$.

While being defined in the holomorphic representation, the
costratified Hilbert space structure may be transferred to the
Schr\"odinger representation. In Section \rref{SSU2} we shall
determine the costratified Hilbert space structure explicitly for
the case $\group=\mathrm{SU}(2)$.


\subsection{Observables}


The prequantization procedure assigns to a classical observable
$f\in C^\infty(\ctg\group)$ the operator $\hat f$ on the
prequantum Hilbert space $L^2(\ctg\group,\ve)$ given by
\begin{equation}
\hat f = \mr i \hbar X_f +  f-\frac 1 {\hbar} \theta(X_f)\,;
\label{prequant} \end{equation} here $\theta$ is the symplectic
potential \eqref{GsplPot}, so that $-d\theta$ coincides with the
cotangent bundle symplectic structure $\omega$ on $\ctg\group$,
and  $X_f$ denotes the Hamiltonian vector field associated with
$f$, determined by the identity
$$
\omega(X_f,\,\cdot\,) = df,
$$
in accordance with Hamilton's equations. The formula
\eqref{prequant} is essentially the same as that given as (8.2.2)
in \cite{woodhous}, save that the Hamiltonian vector field $X_f$
and the symplectic potential $\theta$ are the negatives of the
corresponding objects in \cite{woodhous}. Let $\{\,\cdot \, ,
\,\cdot \,\}$ be the Poisson structure on $C^\infty(\ctg\group)$
associated with the cotangent bundle symplectic structure
$\omega$; this Poisson structure is given by \eqref{GPoibra}. Then $\hbar \{\,\cdot \,
, \,\cdot \,\}$ is the Poisson structure on $C^\infty(\ctg\group)$
associated with the symplectic structure $\frac{\omega}{\hbar}$.
The formula \eqref{prequant} yields a representation of the Lie
algebra underlying the Poisson algebra
$\left(C^\infty(\ctg\group),\hbar \{\,\cdot \, , \,\cdot
\,\}\right)$ which satisfies the Dirac conditions. This
representation is not irreducible and, to arrive at an irreducible
representation of at least a certain subalgebra, the standard
procedure is to introduce a polarization. Observables in this
subalgebra are then referred to as being quantizable in the
polarization under discussion.

In our situation, in the K\"ahler polarization, only the constants
are quantizable. In the Schr\"odinger polarization, the
topological obstruction to the existence of a half-form bundle
vanishes for trivial reasons and, with the half-form correction
incorporated, the relevant subalgebra of $C^\infty(\ctg\group)$
contains the functions which restrict to polynomials of at most
second order on the fibres of $\ctg\group$, i.~e., which are at
most quadratic in the generalized momenta. Thus, it contains the
(classical) Hamiltonian \eqref{GHaFn} of our model. The associated
quantum observable, i.~e., the (quantum) Hamiltonian, is given by
 \beq\label{GHaOp}
H = -\frac{\hbar^2}{2}\Delta_\group + \frac{\inco}{2}
(3-\Re\chi_{\lambda_1})\,,
 \eeq
where $\lambda_1$ denotes the highest weight of the defining
representation of $\group$. The operator $\Delta_\group$ arises
from the non-positive Laplace-Beltrami operator
$\tilde\Delta_\group$ associated with the bi-invariant Riemannian
metric on $\group$ as follows: The operator $\tilde\Delta_\group$
is essentially self-adjoint on $C^\infty(\group)$ and has a unique
extension $\Delta_\group$ to an (unbounded) self-adjoint operator
on $L^2(\group,dx)$. The spectrum being discrete, the domain of
this extensions is the space of functions of the form $f=\sum_n
\alpha_n \vp_n$ such that $\sum_n |\alpha_n|^2 \lambda_n^2  <
\infty$ where the $\vp_n$'s range over the eigenfunctions and the
$\lambda_n$'s over the eigenvalues of $\tilde\Delta_\group$.

Since the metric is bi-invariant, so is the operator
$\Delta_\group$, whence this operator restricts to a self-adjoint
operator on the subspace $L^2(\group,\mr d x)^\group$ which we
continue to denote by $\Delta_\group$. A core for this operator,
and hence for the Hamiltonian $H$, is given by
$C^\infty(\group)^\group$.

By means of the unitary transform \eqref{H} we now transfer the
Hamiltonian and, in particular, the operator $\Delta_\group$ to
the holomorphic representation, i.~e., to self-adjoint operators
on $\mc HL^2(\group^\CC,\mr e^{-\kappa/\hbar}\eta\ve)^\group$.
Concerning $\Delta_\group$, we may alternatively view
$\tilde\Delta_\group$ as a differential operator on
$\group^{\mathbb C}$ via the embedding of $\lieal$ into
$\lieal^{\mathbb C}$, extend it to a self-adjoint operator on
$\mathcal H L^2(\group^{\mathbb C},\mathrm e^{-\kappa/\hbar} \eta
\varepsilon)$, and take the restriction to the subspace $\mathcal
H L^2(\group^{\mathbb C},\mathrm e^{-\kappa/\hbar} \eta
\varepsilon)^\group$.

Next, we determine the eigenvalues and the eigenfunctions of
$\Delta_\group$. The operator $\tilde\Delta_\group$ is known to
coincide with the Casimir operator on $\group$ associated
with the bi-invariant Riemannian metric, see \cite{taylothr} (A
1.2). That is to say, after a choice $X_1,\dots,X_m$ of
orthonormal basis of $\lieal$ has been made,
$$
\tilde\Delta_\group = X^2_1 +\dots + X^2_m
$$
in the universal enveloping algebra $\mathrm U(\lieal)$ of $\lieal$, cf.
e.~g. \cite{enelson} (p.~591). Since $\tilde\Delta_\group$ is
bi-invariant, by Schur's lemma, each isotypical $(\group\times
\group)$-summand $L^2(\group,dx)_\lambda$ of $L^2(\group,dx)$ in
the Peter-Weyl decomposition is an eigenspace, and the
representative functions are eigenfunctions for
$\tilde\Delta_\group$. The eigenvalue of $\tilde\Delta_\group$
corresponding to the highest weight $\lambda$ is known to be given
explicitly by $-\varepsilon_{\lambda}$ where
 \begin{equation}
\varepsilon_{\lambda}=(|\lambda+\rho|^2-|\rho|^2),
 \label{EWL}
 \end{equation}
cf. e.~g.\ \cite{helgaboo} (Chap. V.1 (16)). The sign is chosen in such a way 
that the $\ve_\lambda$ can be interpreted as energy values. Hence,
in particular, each character $\chi_\lambda$ is an eigenfunction
of $\Delta_\group$ associated with the eigenvalue $-\ve_\lambda$.
Consequently, $\Delta_\group$ being viewed as an operator on the
abstract Hilbert space $\Hi$, the vectors $\ket\lambda \in \Hi$
form an orthonormal eigenbasis of  $\Hi$. In view of an
observation spelled out above, the domain of $\Delta_\group$ is
explicitly given by
 \beq\label{Gdomain}
 \left\{
\sum\nolimits_\lambda \alpha_\lambda \ket \lambda \in \Hi
 :
\sum\nolimits_\lambda |\alpha_\lambda|^2 \ve_\lambda^2 < \infty
 \right\}\,.
 \eeq


\section{The costratified Hilbert space for $\SU(2)$}
\label{SSU2}



\subsection{Group theoretical data}


The (real) root system of $\lieal = \su(2)$ consists of the two
roots $\alpha$ and $-\alpha$, given by
$$
\alpha\big(Y) = 2 y,
 \,~~~~~~
Y\in\mf t\,,
$$
where $Y=\diag(\mr i y ,- \mr i y)$, $y\in\RR$. Then $\halfsum =
\frac12\alpha$. We label the irreducible representations
by non-negative integers $n$ (twice the spin). The corresponding
highest weights $\lambda_n$ are given by $\lambda_n = \frac n 2
\alpha$. On $T\times\mf t \cong T^\CC$, the corresponding complex characters
$\chi^\CC_n$ of $\group^\CC = \SL(2,\CC)$ are given by
 \beq\label{GchiC}
\chi^\CC_n(t) = z^n + z^{n-2} + \cdots + z^{-n}
 \,,~~~~~~
t\in T^\CC\,,
 \eeq
where $t = \diag(z,z^{-1})$, $z\in \CC^\ast$. Restriction to
$\group$ yields the real characters which, on $T$, can be written
as
 \beq\label{Gchi}
\chi_n(t) = \frac{\sin\big((n+1)x\big)}{\sin(x)}
 \,,~~~~~~
t\in T\,,
 \eeq
where $t=\diag(\mr e^{\mr i x}, \mr e^{- \mr i x})$, $x\in\RR$.
Any invariant inner product  $\langle \,\cdot \, , \,\cdot \,
\rangle$ on $\lieal = \su(2)$ is proportional to the (negative
definite) trace form. Hence, given $\langle \,\cdot \,  ,\,\cdot
\, \rangle$, we can define a positive number $\scale$ by
$$
\langle Y_1,Y_2 \rangle = -\frac{1}{2\scale^2} \tr(Y_1Y_2),\
Y_1,Y_2 \in \lieal\,.
$$
For the Killing form, $\scale = \frac 1 {\sqrt 8}$. Relative to the given 
invariant inner product on $\lieal$, the two roots
$\alpha$ and $-\alpha$ have norm $|\alpha|^2 = 4 \scale^2$. Hence
$|\halfsum|^2 = \scale^2$ and $|\lambda_n+\halfsum|^2 =
\scale^2(n+1)^2$ whence according to \eqref{G-Cl} and \eqref{EWL}
 \beq\label{GEWCn}
\ve_n = \scale^2 n (n+2)
 \,,~~~~~~
C_n = (\hbar\pi)^{3/2} \mr e^{\hbar\scale^2(n+1)^2}\,.
 \eeq


\subsection{The costratified Hilbert space structure}
\label{costratified}

According to Section \rref{Squantum}, the appropriate Hilbert
space for the holomorphic representation is the Hilbert space $\mc H
L^2(\group^\CC,\mr e^{-\kappa/\hbar} \eta\ve)^\group$ or,
equivalently, $\mc H L^2(T^\CC,\mr e^{-\kappa/\hbar} \gamma
\ve_T)^W$. The Hilbert space for the Schr\"odinger representation is
the space $L^2(\group,\mr d x)^\group$ or, equivalently,
$L^2(T,v\,\mr d t)^W$. There is no need to spell out the functions
$\kappa$, $\eta$, $\gamma$ or $v$ here, because we can work
entirely in the basis given by the characters. For $n \geq 0$, let
$|n\rangle := \ket{\lambda_n}$; then $\{\ket n : n=0,1,2,\dots\}$
is an orthonormal basis of $\Hi$, and we can pass to the
holomorphic and to the Schr\"odinger representation by replacing
each $\ket n$ with the corresponding (normalized) character.

We now determine the costratified Hilbert space structure
constituents $\Hi_\pm$ and $\Hi_\prin$ associated with the strata
$\pha_\pm$ and $\pha_1$ of $\pha$ and the subspace $\Hi_\sing$
associated with the orbit type subset $\pha_0$. Recall the
notation and the description of these strata from Subsection
\rref{SSsymred}. As $\pha_\prin$ is the top stratum, $\Hi_\prin =
\Hi$. To describe the subspaces $\Hi_\pm$ and $\Hi_\sing$, we pass
to the holomorphic representation.

\ble\label{Lvani}

The systems \eqref{Gvanipbs1}, \eqref{Gvanimbs1}, and \eqref{Gvanisingbs1} below
constitute bases of, respectively, the subspaces $\vani_+, \vani_-, \vani_\sing$ of $\Hi$
corresponding to, respectively, the strata $\pha_+$, $\pha_-$ and $\pha_\sing:$
 \beqa\label{Gvanipbs1}
\chi^\CC_n - (n+1) \chi^\CC_0
 \,,&~~~~~~&
n = 1,2,3,\dots\,,
\\ \label{Gvanimbs1}
\chi^\CC_n + (-1)^n \frac{n+1}{2} \chi^\CC_1
 \,,&~~~~~~&
n = 0,2,3,\dots\,,
\\ \label{Gvanisingbs1}
\chi^\CC_{2k} - (2k+1) \chi^\CC_0
 \,,~~~~~~
\chi^\CC_{2k+1} - (k+1) \chi^\CC_1
 \,,&~~~~~~&
k = 1,2,3,\dots\,.
 \eeqa

\ele

\begin{proof}

We view the elements of $\Hi$ as functions on $T^\CC$ rather than
on $\group^\CC$. Via the polar decomposition map $T\times\mf t \to
T^\CC$, the points $(\pm\II,0)$ are mapped to $\{\pm\II\}$. Hence,
$\vani_+$, $\vani_-$ and $\vani_\sing$ consist of the functions
$\psi\in\Hi$ that satisfy, respectively,
 \beq\label{Gvanicond}
\psi(\II) = 0
 \,,~~~~~~
\psi(-\II) = 0
 \,,~~~~~~
\psi(\pm\II) = 0
 \,.
 \eeq
Due to $\chi^\CC_n(\pm\II) = (\pm 1)^n(n+1)$, we
have
$$
\chi^\CC_{2k}(\pm\II) = 2k+1 = (2k+1)\chi^\CC_0(\pm\II)
$$
and
$$
\chi^\CC_{2k+1}(\pm\II) = \pm (2k+2) = (k+1)\chi^\CC_1(\pm\II)\,.
$$
Hence, all the functions given in
\eqref{Gvanipbs1}--\eqref{Gvanisingbs1} satisfy the corresponding
condition in \eqref{Gvanicond}. Conversely, given $\psi\in
\vani_+$, expanding it in the basis of $\Hi$ given by the elements
in \eqref{Gvanipbs1} together with $\chi^\CC_0$ we see that the vanishing of
$\psi(\II)$
implies that the coefficient of $\chi^\CC_0$ is zero. The
reasoning for $\vani_-$ and $\vani_\sing$ is analogous. Finally,
linear independence of the systems
\eqref{Gvanipbs1}--\eqref{Gvanisingbs1} is obvious.
\end{proof}

\noindent
We express the bases \eqref{Gvanipbs1}--\eqref{Gvanisingbs1}, up
to a common factor $(\hbar\pi)^{3/4}$, in terms of $\ket n$:
 \beqa\label{Gvanipbs}
\mr e^{\hbar\scale^2(n+1)^2/2} \ket n
 -
(n+1) \mr e^{\hbar\scale^2/2} \ket 0
 \,,&&
n = 1,2,3,\dots\,,
\\ \label{Gvanimbs}
\mr e^{\hbar\scale^2(n+1)^2/2} \ket n
 -
\frac{n+1}{2} \mr e^{2 \hbar\scale^2} \ket 1
 \,,&&
n = 0,2,3,\dots\,,
\\ \nonumber
\mr e^{\hbar\scale^2(2k+1)^2/2} \ket {2k}
 -
(2k+1) \mr e^{\hbar\scale^2/2} \ket{0}
 \,, && k = 1,2,3,\dots\,,
\\ \label{Gvanisingbs}
\mr e^{2\hbar\scale^2(k+1)^2} \ket {2k+1}
 -
(k+1) \mr e^{2\hbar\scale^2} \ket 1
 \,,&&
k = 1,2,3,\dots\,.
 \eeqa

\bpp\label{PHsing}

The subspaces $\Hi_+$ and $\Hi_-$ have dimension $1$. They are
spanned by the normalized vectors
 \beqa \label{Gpsip}
\psi_+ & := & \frac{1}{N} \sum\nolimits_{n=0}^\infty (n+1)\, \mr
e^{-\hbar\scale^2\,(n+1)^2/2}\, \ket n
 \,,
\\ \label{Gpsim}
\psi_- & := & \frac{1}{N} \sum\nolimits_{n=0}^\infty (-1)^n\, (n+1) \, \mr
e^{-\hbar\scale^2 \, (n+1)^2/2} \, \ket n\,,
 \eeqa
respectively. The subspace $\Hi_\sing$ has dimension $2$. It is
spanned by the orthonormal basis
 \beq\label{Gpsising}
\psi_\even
 :=
 \frac{1}{N_\even} \sum_{\text{\rm $n$ even}}
(n+1) \, \mr e^{-\hbar\scale^2\,(n+1)^2/2} \ket{n}
 \,,~~~~
\psi_\odd
 :=
 \frac{1}{N_\odd} \sum_{\text{\rm $n$ odd}}
(n+1) \, \mr e^{-\hbar\scale^2\,(n+1)^2/2} \ket{n}\,,
 \eeq
where the sum over the even $n$ includes $n=0$. The normalization factors are
$$
N^2 = \sum_{n=1}^\infty n^2 \, \mr e^{-\hbar\scale^2 \, n^2}
 \,,~~~~~~
N_\even^2 = \sum_{\text{\rm $n$ odd}} n^2 \mr e^{-\hbar\scale^2 \,
n^2}
 \,,~~~~~~
N_\odd^2 = \sum_{\text{\rm $n$ even}} n^2 \mr e^{-\hbar\scale^2 \,
n^2}\,.
$$

\epp

\bpf

The sums in \eqref{Gpsip}, \eqref{Gpsim} and \eqref{Gpsising}
converge, their limits are normalized, and $\psi_\even$ and
$\psi_\odd$ are mutually orthogonal. The vector $\psi_+$ together
with the system \eqref{Gvanipbs}, $\psi_-$ together with the
system \eqref{Gvanimbs}, and $\psi_\even$, $\psi_\odd$ together
with the system \eqref{Gvanisingbs} provide bases of $\Hi$.
Finally, it is straightforward to check that $\psi_+$, $\psi_-$
and $\psi_\even$, $\psi_\odd$ are orthogonal to the corresponding
system in \eqref{Gvanipbs}--\eqref{Gvanisingbs}. \epf

\noindent
Proposition \rref{PHsing} implies that, in Dirac notation, for $i=\sing,\pm$,
the orthogonal projections $\Pi_i\equiv\Pi_{\prin i}\colon\Hi_\prin\to\Hi_i$
are given by
 \beq\label{Goproj}
\Pi_\pm
 =
|\psi_\pm\rangle\,\langle\psi_\pm|
 \,,~~~~~~
\Pi_\sing
 =
|\psi_\even\rangle\,\langle\psi_\even|
 +
|\psi_\odd\rangle\,\langle\psi_\odd|
 \,.
 \eeq
The normalization factors $N$, $N_\even$ and $N_\odd$ can be
expressed in terms of the $\theta$-constant $\theta_3$ with \lq nome\rq\  $Q$ as
 \beq\label{Gdeftheta}
\theta_3(Q) = \sum_{k=-\infty}^\infty Q^{k^2}\,.
 \eeq
For example,
 \beq\label{Gnormtheta}
N^2
 =
- \frac{\partial}{\partial (\hbar\scale^2)}
\sum_{n=1}^\infty \mr e^{-\hbar\scale^2 \, n^2}
 =
- \frac{1}{2} \frac{\partial}{\partial (\hbar\scale^2)}
\theta_3(\mr e^{-\hbar\scale^2})
 =
\frac 1 2 \mr e^{-\hbar\scale^2} \theta_3'(\mr e^{-\hbar\scale^2})\,.
 \eeq
Then
 \beq\label{Gnorm01theta}
N_\odd^2
 =
4 \mr e^{-4 \hbar\scale^2}\theta_3'(\mr e^{-4
\hbar\scale^2})
 \,,~~~~~~
N_\even^2
 =
N^2 - N_\odd^2\,.
 \eeq

\bre \label{R-cs}

Let $\rho_\hbar(a) := \mr e^{\hbar\beta^2} N \psi_+$. Using \eqref{Gpsip} and
plugging in for $\ket n$ the real characters $\chi_n$ we see that $\rho_\hbar$
satisfies the heat equation $\frac{\mr d}{\mr 
d\hbar} \rho = \frac12\Delta_\group\rho_\hbar$ subject to the initial condition 
$\rho_0 = \delta_\II$, i.e., $\rho_t$ is the heat kernel of $\group$. The
expansion of $\rho_\hbar$ obtained from \eqref{Gpsip} is the standard expansion
of the heat kernel of a compact Lie group in terms of its characters \cite[p.\
38]{Stein}. According to \cite[\S 4, Prop.\ 1]{Hall94}, the function
$\rho_\hbar$ has an analytic continuation to $K^\CC$. This analytic continuation
does not consist in substitution of the character $\chi_n^\CC$ for the character
$\chi_n$ in  the standard expansion; in particular, the resulting formal series
does not converge in $\mc H L^2(\group^\CC,\mr e^{-\kappa/\hbar}
\eta\ve)^\group$. Thus $\rho_\hbar$ defines the complex-valued functions  
$$
\psi^{(\hbar)}_g(a) = \ol{\rho_\hbar(ga^{-1})}
 \,,~~~~~~
a\in \group\,,
$$
on $\group$, parametrized by the members $g\in\group^\CC$. 
According to \cite{Hall94}, these functions admit an interpretation as coherent
states on $\group$. Indeed, the functions $\psi_\pm $ and
$\psi^{(\hbar)}_{\pm\II}$ are related by the identity
$$
\psi_\pm = \frac{\mr e^{-\hbar\beta^2}}{N} \psi^{(\hbar)}_{\pm\II}\,,
$$
i.e., up to a normalization factor, the states spanning the subspaces 
$\Hi_\pm$
are the coherent states labelled by the points of the corresponding 
strata. This observation is certainly not a coincidence; in fact, for physical
reasons, the states which the functions $\psi_\pm $ represent
should come down to coherent states because the (phase space) wave function
orthogonal to all wave functions vanishing at a given point represents a state
of optimal localization in phase space (i.e., minimal position-momentum
uncertainty). This is exactly what is generally understood to be  a coherent
state. 

\ere


\subsection{Tunneling between strata}


Consider the constituents $\Hi_+$ and $\Hi_-$ of the
costratified Hilbert space $\Hi$ relative to the orbit type stratification
of $\pha$. A straightforward calculation yields
$$
\langle \psi_+,\psi_-\rangle
 =
 \frac{1}{N^2} \sum_{n=1}^\infty (-1)^{n+1} \,
n^2 \, \mr e^{-\hbar\scale^2\,n^2}
 =
\frac{N_\even^2 - N_\odd^2}{N^2}\,.
$$
As in subsection \ref{costratified} above, the scalar product can
be expressed in terms of $\theta$-functions. Likewise, as in
\eqref{Gnormtheta} for $N^2$, the alternating sum in the
denominator can be rewritten as
 $
- \mr e^{-\hbar\scale^2}\theta_3'(-\mr e^{-\hbar\scale^2})\,.
 $
Together with \eqref{Gnormtheta} this yields
 \beq\label{G-scapro-theta}
\langle \psi_+,\psi_-\rangle
 =
 -\, \frac{
\theta_3'\big(-\mr e^{-\hbar\scale^2}\big)
 }{
\theta_3'\big(\mr e^{-\hbar\scale^2}\big)
 }\,.
 \eeq
The absolute square $|\langle \psi_+,\psi_-\rangle|^2$ is the
tunneling probability between the strata $\pha_+$ and $\pha_-$,
i.~e., the probability for a state prepared at $\pha_+$ to be
measured at $\pha_-$ and vice versa.

\begin{figure}

\begin{center}

\hspace*{-2.5cm}\epsfig{file=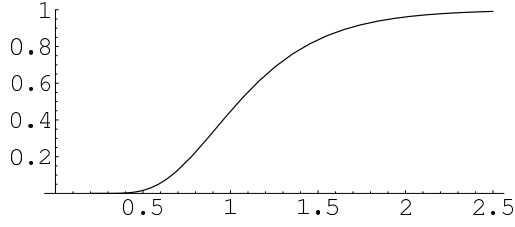,width=10cm}

\caption{\label{AbbTunnelW} Tunneling probability $|\langle
\psi_+,\psi_-\rangle|^2$ as a function of $\hbar\scale^2$}

\end{center}

 \end{figure}

The numerical value of this quantity strongly depends on the
combined constant $\hbar\scale^2$, see Figure \rref{AbbTunnelW}.
For large values of $\hbar\scale^2$, $|\langle \psi_+,\psi_-\rangle|^2$
is almost equal to $1$. This can
also be read off from the expansions \eqref{Gpsip} and
\eqref{Gpsim}: the first coefficient that distinguishes between
$\psi_+$ and $\psi_-$ is $\mr 2e^{-4\hbar\scale^2}$;  for
large $\hbar\scale^2$, this coefficient
is much smaller than the leading coefficient
$\mr e^{-\hbar\scale^2}$, so that $\psi_+$ and $\psi_-$ have a
large overlap. In fact, in the limit $\hbar\scale^2\to\infty$ they become both
equal to $\ket 0$.

On the other hand, for $\hbar\scale^2\to 0$ we have
$|\langle \psi_+,\psi_-\rangle|^2 \to 0$. Thus, in the semiclassical limit,
 the tunneling probability
vanishes.

\bre

Since the strata $\pha_+$ and $\pha_-$ together constitute
the orbit type subset
$\pha_\sing$, a tunneling between them should not be
visible in the costratification given by $\Hi_\sing$, that is, in
the costratification
relative to the coarser decomposition $\pha=\pha_\sing\cup\pha_\prin$ by mere 
orbit types (and not by the connected components thereof). Indeed, we have
 \beq\label{GHsingdeco}
\Hi_\sing = \Hi_+ \oplus \Hi_-\,,
 \eeq
where the sum is direct but not orthogonal, and $\psi_\pm$ can be written as
$$
\psi_\pm = \frac{N_\even}{N} \psi_\even \pm \frac{N_\odd}{N} \psi_\odd\,.
$$
In other words, the subspaces
$\Hi_+$ and $\Hi_-$ are swallowed by $\Hi_\sing$
and there is no way to reconstruct them from $\Hi_\sing$
alone.

\ere

\bre

Expressing the scalar product in terms of the coherent states
$\psi^{(\hbar)}_\II$ and $\psi^{(\hbar)}_{-\II}$, see Remark \rref{R-cs},
we obtain the identity 
$$
|\langle \psi_+,\psi_- \rangle|^2
 =
 \frac{
\big|\big\langle \psi^{(\hbar)}_{\II},\psi^{(\hbar)}_{-\II} \big\rangle\big|^2
 }{
\|\psi^{(\hbar)}_{\II}\|^2\,\|\psi^{(\hbar)}_{-\II}\|^2
 }\,.
$$
The quantity
 $ 
 \frac{
\big|\big\langle \psi^{(\hbar)}_g,\psi^{(\hbar)}_h \big\rangle\big|^2
 }{
\|\psi^{(\hbar)}_g\|^2\,\|\psi^{(\hbar)}_h\|^2
 }
 $
is known as the overlap of the coherent states $\psi^{(\hbar)}_g$ and
$\psi^{(\hbar)}_h$; it was studied in more general situations in great detail in
a series of papers by Thiemann and collaborators
\cite{Thiemann,ThiemannWinkler}. Among other 
things, they have shown that for $K=\SU(2)$ the overlap is related with the
geodesic distance on $\group^\CC = \SL(2,\CC)$ and that, in general, 
for $g\neq h$ and $\hbar\to 0$, the overlap vanishes faster than any power of
$\hbar$. 

The scalar product $\langle \psi^{(\hbar)}_g,\psi^{(\hbar)}_h \rangle$,
viewed as a function of $g$ and $h$, is known as the reproducing kernel
associated with the familiy of coherent states $\psi^{(\hbar)}_g$,
$g\in\group^\CC$. It can be expressed in terms of the heat 
kernel $\rho_\hbar$. For $\SU(2)$, this leads to Formula
\eqref{G-scapro-theta}. 

\ere


\subsection{Adapted orthonormal bases}


For $i=\pm,\sing$, we will now construct orthonormal bases of the subspaces
$\vani_i$ of $\Hi$. To this end, let
$$
\hat\psi_\pm
 :=
\frac{(1-\Pi_\mp)\psi_\pm}{\|(1-\Pi_\mp)\psi_\pm\|} .
$$
Then $\vani_\pm = \vani_\sing \oplus \CC\hat\psi_\mp$,
the sum being orthogonal since $\hat\psi_\pm\in\Hi_\sing$.
Hence, it suffices to construct an orthonormal basis of $\vani_\sing$.
For that purpose, we orthonormalize the family \eqref{Gvanisingbs}. This can
of course be done for the even and odd degree families separately.

\begin{Lemma}\label{Logon}

Let $\vp_n$, $n=0,1,2,\dots$ be an orthonormal basis of the Hilbert
space $\mc E$ and let $\vanifac_n$, $n=0,1,2,\dots$, be real numbers with
$\vanifac_0=1$. Then orthonormalization of the system $\vp_n -
\vanifac_n \vp_0$, $n=1,2,3,\dots$,  yields the system
 \beq\label{GvaniONB}
\tilde\vp_n
 ~=~
\frac{\vanisum_{n-1}}{\vanisum_n} \vp_n ~-~
\frac{\vanifac_n}{\vanisum_n \vanisum_{n-1}} \sum_{k=0}^{n-1}
\vanifac_k\,\vp_k
 \,,~~~~~~
n=1,2,3,\dots\,,
 \eeq
where $\vanisum_n^2 = \sum_{k=0}^n \vanifac_k^2$.

\end{Lemma}

\begin{proof}

Straightforward calculation.
\end{proof}

Let $\psi_{2n}$, $n=1,2,3,\dots$, denote the basis elements
obtained by application of Lemma \rref{Logon} to the even degree family of
\eqref{Gvanisingbs}. Thus substituting
$\ket{2k}$ for $\vp_k$ and $(2k+1) \mr e^{-\hbar\ve_{2k}/2}$
for $\vanifac_k$ in \eqref{GvaniONB} yields $\psi_{2n}$.
Likewise, let $\psi_{2n+1}$, $n=1,2,3,\dots$, denote the
basis elements obtained by applying the lemma to the odd degree
family of \eqref{Gvanisingbs}, so that substituting
$\ket{2k+1}$ for $\vp_k$ and
$(k+1) \mr e^{-2\hbar\ve_k}$ for $\vanifac_k$ in
\eqref{GvaniONB} yields
 $\psi_{2n+1}$.

The resulting vectors $\psi_n$, $n=2,3,4,\dots$ form an orthonormal basis of
$\vani_\sing$. Adding $\hat\psi_-$, we obtain an orthonormal basis of
$\vani_+$. Adding $\hat\psi_+$, we obtain an orthonormal basis of
$\vani_-$.


\subsection{Representation in terms of  $L^2[0,\pi]$}


From now on we will work in the Schr\"odinger representation,
i.e., we realize $\Hi$ as
$$
L^2(\group)^\group \cong L^2(T,v\mr d
t)^W.
$$
In order to produce plots of wave functions $\psi\in\Hi$ we choose
a suitable parameterization of $\cfg$ and represent the elements
of $\Hi$ by ordinary $L^2$-integrable functions on the parameter
space. This representation will also be used in the discussion of
the stationary Schr\"odinger equation of our model in Section
\rref{Seigen}. A suitable parameterization of $\cfg$ can be
obtained as follows. We parameterize $T$ by
 \beq\label{GparamT}
\diag(\mr e^{\mr i x},\mr e^{-\mr i x})
 \,,~~~~~~
x\in[-\pi,\pi]\,.
 \eeq
Since the nontrivial element of $W$ acts by reflection $x\mapsto
-x$, restriction of the parameter $x$ to the interval $[0,\pi]$
yields a (bijective) parameterization of $\cfg$, where $\cfg_+$
corresponds to $x=0$ and $\cfg_-$ to $x=\pi$.

In the parameterization \eqref{GparamT}, the measure $v\,\mr d t$
on $T$ is given by
$$
v\,\mr d t = \frac{\vol(\group)}{\pi} \sin^2(x) \,\mr d x \,.
$$
Hence, the assignment to $\psi\in C^\infty(T)^W$ of the function
$x\mapsto \psi(\diag(\mr e^{\mr i x},\mr e^{-\mr i x}))$,
$x\in[0,\pi]$ defines a Hilbert space isomorphism
 \beq\label{GGamma1}
\Gamma_1\colon \Hi \to L^2([0,\pi],\sin^2(x) \mr d x)\,.
 \eeq
Furthermore, multiplication by $\sqrt 2 \sin x$ defines a Hilbert
space isomorphism
 \beq\label{GGamma2}
\Gamma_2\colon L^2([0,\pi],\sin^2(x) \mr d x) \to L^2[0,\pi]\,.
 \eeq
Here the scalar products in $L^2([0,\pi],\sin^2(x) \mr d x)$ and
$L^2[0,\pi]$ are normalized so that the constant function with
value $1$ has norm $1$.

The composite isomorphism $\Gamma = \Gamma_2\circ\Gamma_1$
identifies $\Hi$ with the space $L^2[0,\pi]$ of ordinary
square-integrable functions on $[0,\pi]$. Plotting the 
function $\Gamma\psi$ rather than $\psi$ has the advantage that
one can read off directly from the graph the corresponding
probability density with respect to Lebesgue measure on the
parameter space $[0,\pi]$.

\begin{figure}
 \begin{center}

\unitlength1cm

 \begin{picture}(12,2.5)
 \put(0.1,0){
 \marke{0,0}{cc}{\psi_+}
 \marke{4,0}{cc}{\psi_-}
 \marke{8,0}{cc}{\psi_\even}
 \marke{12,0}{cc}{\psi_\odd}
 }
 \put(-2.8,-0.2){
 \put(0,0){\epsfig{file=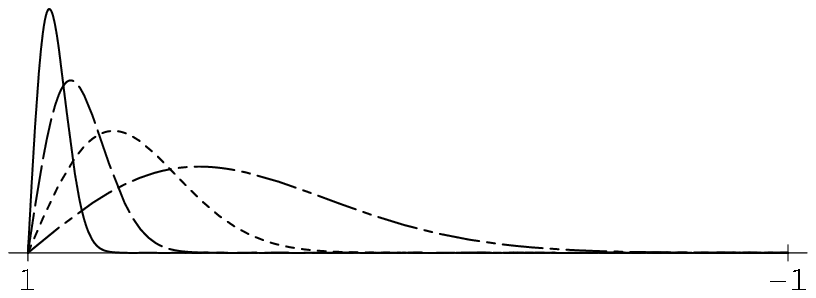,width=4.5cm,height=2.5cm}}
 \put(4,0){\epsfig{file=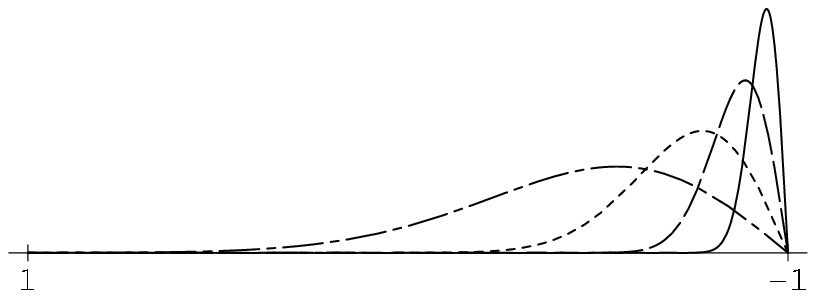,width=4.5cm,height=2.5cm}}
 \put(8,-0.2){\epsfig{file=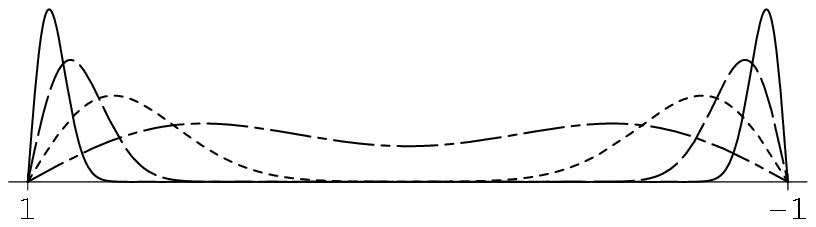,width=4.5cm,height=2.5cm}}
 \put(12,-0.2){\epsfig{file=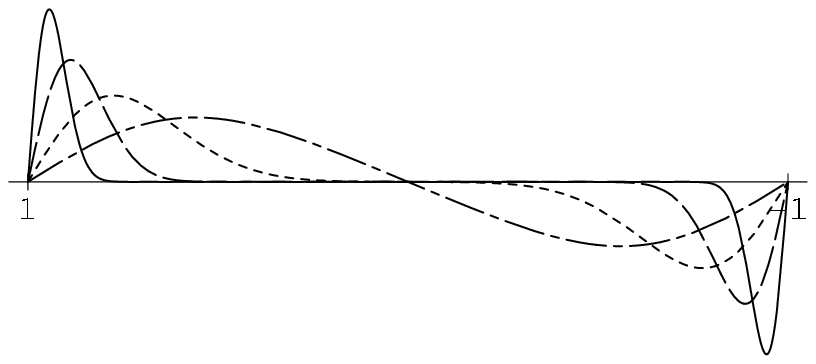,width=4.5cm,height=2.5cm}}
  }
 \end{picture}

 \end{center}

\caption{\label{Abbpsipm} Plots of images of the wave functions $\psi_i$,
$i=\pm,\even,\odd$ under $\Gamma$, for $\hbar\scale^2 =
1/128$ (continuous line), $1/32$ (long dash), $1/8$ (short dash),
$1/2$ (alternating short-long dash).}

 \end{figure}

Plots of $\Gamma\psi_i$, $i=\pm,\even,\odd$, are shown in Figure
\rref{Abbpsipm} for $\hbar\scale^2 = \frac 1 2, \frac 1 8, \frac 1
{32}, \frac 1 {128}$. We remark that the value $\hbar\scale^2 =
1/8$ appears when we choose $\hbar =1$ and the negative of the
Killing form as the invariant scalar product on $\mf g$. Moreover,
according to \eqref{Gchi},
 \beq\label{GGammachi}
\big(\Gamma\chi_n\big)(x) = \sqrt 2 \sin\big((n+1)x\big)\,,
 \eeq
hence the expansions \eqref{Gpsip}--\eqref{Gpsising} boil down to
ordinary Fourier expansions of the functions $\Gamma\psi_i$,
$i=\pm,\even,\odd$.

\begin{figure}
 \begin{center}

\unitlength1cm

 \begin{picture}(12,5.5)
\put(0,0.7){
 \put(0.1,2.5){
 \marke{0,0}{cc}{\psi_2}
 \marke{4,0}{cc}{\psi_3}
 \marke{8,0}{cc}{\psi_4}
 \marke{12,0}{cc}{\psi_5}
 }
 \put(-2.8,3){
 \put(0,0){\epsfig{file=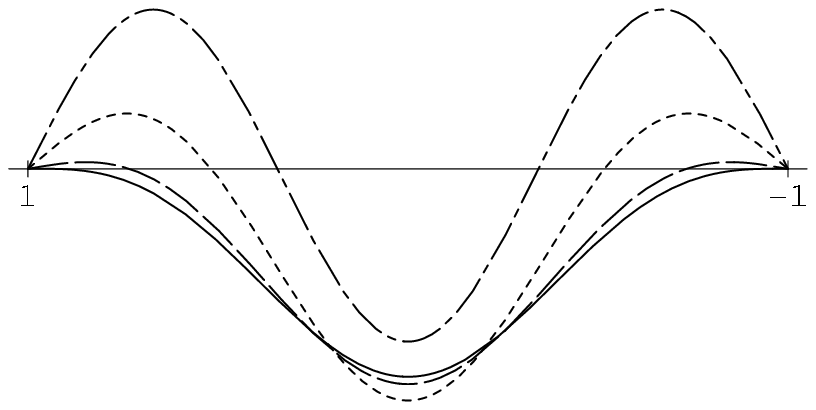,width=4.5cm}}
 \put(4,0){\epsfig{file=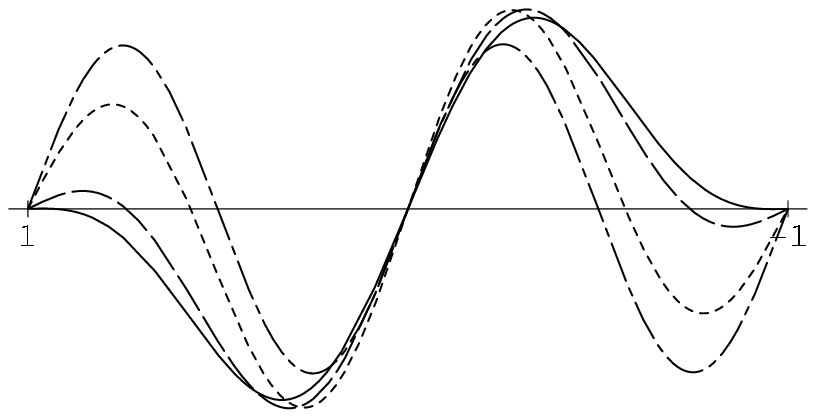,width=4.5cm}}
 \put(8,-0.2){\epsfig{file=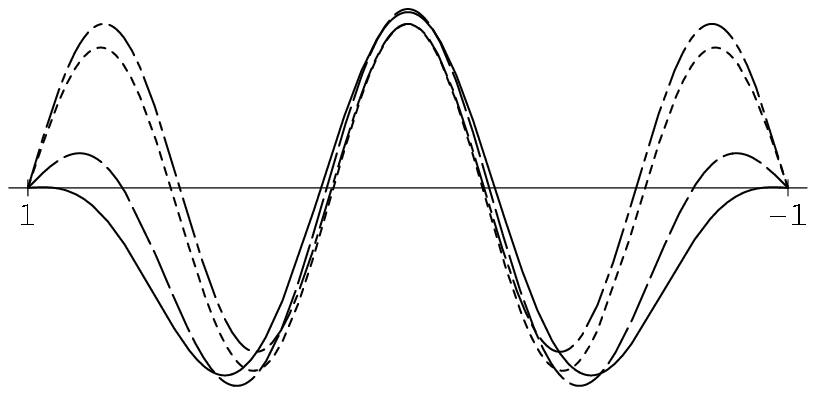,width=4.5cm}}
 \put(12,-0.2){\epsfig{file=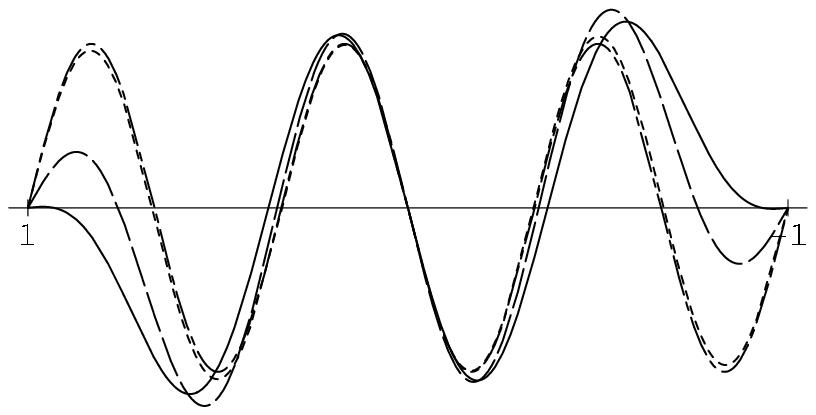,width=4.5cm}}
  }
 \put(0.1,-0.7){
 \marke{2.5,0}{cc}{\hat\psi_+}
 \marke{9.5,0}{cc}{\hat\psi_-}
 }
 \put(-2.2,-0.2){
 \put(2.5,0){\epsfig{file=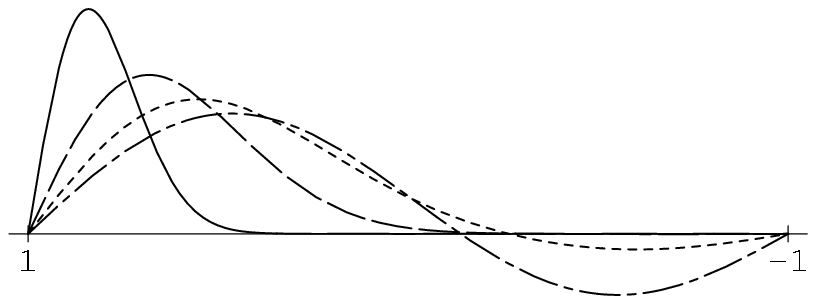,width=4.5cm}}
 \put(9.5,0){\epsfig{file=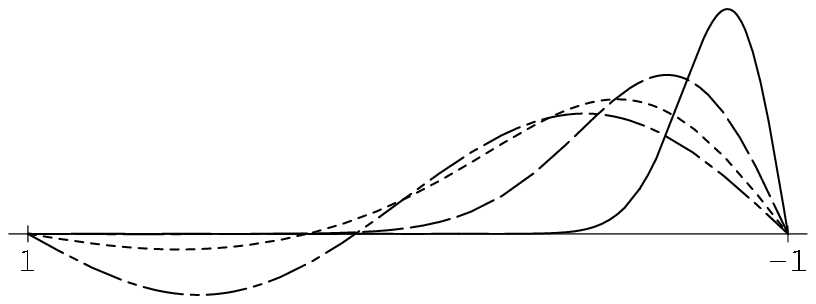,width=4.5cm}}
 }
}
 \end{picture}

 \end{center}

\caption{\label{AbbvaniONB} Plots of the images of the wave functions $\psi_2,
\dots, \psi_4$ and $\hat\psi_\pm$, under $\Gamma$, for
$\hbar\scale^2 = \frac{1}{16}$ (continuous line), $\frac 1 4$
(long dash), $\frac 1 2$ (short dash), $1$ (alternating short-long
dash).}

 \end{figure}

Plots of $\psi_2,\dots,\psi_5$ and $\hat\psi_\pm$ are shown in
Figure \rref{AbbvaniONB} for $\hbar\scale^2 =  1, \frac 1 2, \frac
1 4, \frac{1}{16}$. For $\hbar\scale^2\to 0$, the outer nodes of
the $\Gamma\psi_n$ run into the points $\cfg_\pm$ and thus
decrease the number of nodes to $n-2$. Moreover, since for
decreasing value of $\hbar\scale^2$ the overlap $\langle
\psi_+,\psi_-\rangle$ decreases, the functions $\hat\psi_\pm$
converge to $\psi_\pm$.

\section{Energy eigenvalues and eigenstates for $\SU(2)$}
\label{Seigen}


We now determine the energy eigenvalues and the corresponding
eigenfunctions of our model for $K = \SU(2)$. We start with a
general discussion of the Hamiltonian.


\subsection{The Hamiltonian}
\label{SSHam}


In the Schr\"odinger representation, the Hamiltonian is given by
\eqref{GHaOp}. It is a self-adjoint operator on the Hilbert space
$L^2(\group,\mr d x)^\group \equiv L^2(T,v\mr d t)^W$. For domain
issues it suffices to consider the kinetic part, i.\ e., the
Laplacian $\Delta_\group$. As a core we may take
$C^\infty(\group)^\group \equiv C^\infty(T)^W$. According to
\eqref{Gdomain}, the full domain is
$$
 \left\{
\sum\nolimits_{n=0}^\infty \alpha_n \chi_n \in L^2(\group,\mr d
x)^\group
 :
\sum\nolimits_{n=0}^\infty |\alpha_n|^2 n^2(n+2)^2 < \infty
 \right\}\,.
$$
The isomorphisms $\Gamma_1$ and $\Gamma_2$, see \eqref{GGamma1}
and \eqref{GGamma2}, carry $\Delta_\group$ and $H$ to
the selfadjoint operators
 \begin{align*}
\Delta_1 & = \Gamma_1\circ\Delta_\group\circ\Gamma_1^{-1}
 \,, &
\Delta_2 & = \Gamma_2\circ\Delta_1\circ\Gamma_2^{-1}
 \equiv
\Gamma\circ\Delta_\group\circ\Gamma^{-1}
\\
H_1 & = \Gamma_1\circ H\circ\Gamma_1^{-1}
 \,, &
H_2 & = \Gamma_2\circ H_1\circ\Gamma_2^{-1}
 \equiv
\Gamma\circ H\circ\Gamma^{-1}
 \end{align*}
on the Hilbert spaces $L^2([0,\pi],\sin^2x\mr d x)$ and
$L^2[0,\pi]$, respectively. Then
 \beq\label{GH12}
H_i = -\frac{\hbar^2}{2} \Delta_i + \frac \nu 2 (3-2\cos x)
 \,,~~~~~~
i=1,2\,,
 \eeq
where, formally,
 \beq\label{GDelta12}
\Delta_1
 =
\scale^2 \left(\frac{1}{\sin(x)} \, \frac{\mr d^2}{\mr d x^2}
\sin(x) + 1\right)
 \,,~~~~~~
\Delta_2
 =
\scale^2 \left(\frac{\mr d^2}{\mr d x^2} + 1\right)\,.
 \eeq
The formula for $\Delta_1$ follows from the general formula for
the radial part of the Laplacian on a compact group, see \cite[\S
II.3.4]{helgaboo}, or by explicitly applying this operator to the
functions $\Gamma_1\chi_n$.

Let $C^\infty[0,\pi]$ denote the space of Whitney smooth complex
functions on the closed interval $[0,\pi]$. These are the smooth
functions on the open interval $]0,\pi[$ that can be extended to
smooth functions on $\RR$. In particular, the elements of
$C^\infty[0,\pi]$ have well-defined derivatives of arbitrary order
in $0$ and $\pi$.

\begin{prop}\label{Pcore}

A core for $\Delta_1$ is given by $D_1 = \{ \psi \in
C^\infty[0,\pi] : \psi'(0) = \psi'(\pi) = 0 \}$. A core for
$\Delta_2$ is given by $D_2 = \{ \psi \in C^\infty[0,\pi] :
\psi(0) = \psi(\pi) = 0 \}$.

\end{prop}

\begin{proof}

First, consider $\Delta_1$. We have to show that

(a)~ $\Gamma_1\big(C^\infty(\group)^\group\big) \subseteq D_1$,

(b)~ $\Delta_1(D_1) \subseteq L^2([0,\pi],\sin^2 x \mr d x)$,

(c)~ $\Delta_1$ is symmetric on $D_1$.

\noindent We may replace $\Delta_1$ with the operator
$\tilde\Delta_1 = \frac{1}{\sin x}\,\frac{\mr d^2}{\mr d x^2} \sin
x$. Concerning (a), we observe that the algebra of real invariant
polynomials on $\group = \SU(2)$ is generated by the trace
monomial $\rho(a) = \frac 1 2 \tr(a)$. A theorem in
\cite{Schwarz} states that $C^\infty(\group)^\group = \rho^\ast
C^\infty(\RR)$. Hence, for given $\psi\in C^\infty(\group)^\group$
there exists $\vp\in C^\infty(\RR)$ such that $\psi = \vp\circ
\rho$. Then $(\Gamma_1 \psi)(x) =  \vp(\cos x)$ and thus
$(\Gamma_1 \psi)'(x) = - h'(\cos x) \sin x$ vanishes for
$x=0,\pi$.

To check (b), let $\psi\in D_1$. It suffices to show that the
values of the function $(\tilde \Delta_1 \psi)(x)$, $0<x<\pi$,
converge for $x\to0$ and $x\to\pi$. Since $\psi(0)$ and
$\psi''(0)$ exist,
$$
\lim_{x\to 0} \big(\tilde \Delta_1 \psi(x)\big)
 =
\psi''(0) - \psi(0) + 2 \lim_{x\to 0} \frac{\cos x \psi'(x)}{\sin
x}\,.
$$
Since $\lim_{x\to 0} \psi'(x) = 0$, we can apply the rule of
Bernoulli and de l'Hospital. This yields
$$
\lim_{x\to 0} (\tilde\Delta_1\psi(x)) = 3 \psi''(0) - \psi(0)\,.
$$
The reasoning for $x\to\pi$ is analogous.

To prove (c), let $\psi,\vp\in D_1$. Then, omitting the
normalization factor $2/\pi$, we find
 \begin{align*}
 \int_0^\pi
\ol{\psi(x)} \, (\tilde\Delta_1\vp)(x) \,
 \sin^2 x \mr d x
 = &
 \int_0^\pi
\ol{(\tilde\Delta_1\psi)(x)} \, \vp(x) \, \sin^2 x \mr d x
\\
  & +
\sin x \, \psi(x) (\sin x \, \vp(x))'\big|^\pi_0
 -
\sin x \, \vp(x) (\sin x \, \psi(x))'\big|^\pi_0 \,.
 \end{align*}
The boundary terms vanish because $\psi(x)$, $\psi'(x)$, $\vp(x)$
and $\vp'(x)$ exist for $x=0$ and $x=\pi$.

Next, consider $\Delta_2$. We have to check conditions (a)--(c)
with the subscript $1$ replaced with the subscript $2$, 
with $L^2([0,\pi],\sin^2 x\mr d
x)$ instead of $L^2[0,\pi]$, and with $C^\infty(\group)^\group$
instead of $D_1$. Conditions (a) and (b) are trivially satisfied
and the verification of (c) is analogous to that for $\Delta_1$.

\end{proof}

\bre

The operator $\Delta_1$ is discussed in \cite[\S 4]{Wren:Rieffel}
as a specific example of a reduced Laplacian obtained by Rieffel
induction. There, the same core is isolated. In our concrete
situation the proof is much simpler than in the general setting of
\cite{Wren:Rieffel}, though.

\ere

In view of  the proposition, we will now discuss two items. First,
we will relate our system with two standard elementary quantum
mechanical systems. Thereafter, we will make a remark on the
extension problem of the Hamiltonian in a \lq naive\rq\ 
quantization-after-reduction procedure.

The proposition implies that, for $\nu=0$, the Hilbert space
isomorphism $\Gamma$ maps our original system to that of a
particle of mass $m=\frac 1 {2\scale^2}$ moving in a
one-dimensional square potential well of width $\pi$ with infinitely high
walls. Inside the well the energy is shifted by $\scale^2 = \frac 1
{2m}$. For $\nu\neq 0$, the potential inside the square well is further
modified by a cosine. This corresponds to a planar pendulum that
is bound to move in one half of the circle only and is reflected
elastically at the two equilibria. It would be interesting to
clarify the relevance of the subspaces $\Hi_\pm$ in both these systems.

The relationship with the pendulum is in fact more intimate:
Multiplication by the function $\sqrt 2 \sin x$, $x\in[-\pi,\pi]$,
defines a Hilbert space isomorphism from $L^2(T,v\mr d t)\equiv
L^2([-\pi,\pi],\sin^2 x\mr d x)$ onto $L^2(T,\mr d t) \equiv
L^2[-\pi,\pi]$ which maps the subspace $ \Hi$ of $W$-invariants
onto the subspace of odd functions. The  Hamiltonian is
given formally by the same expression as $H_2$. A core for this
operator is given by the odd $2\pi$-periodic
$C^\infty$-functions on $\RR$. Hence, this operator describes a
planar pendulum of mass $m = \frac{1}{2\scale^2}$ and ratio of
gravitational acceleration by length given by
$\frac{\nu}{\hbar^2\scale^2}$ with the constraint that among the
states of the pendulum only the odd ones emerge. Finally,
restriction to $[0,\pi]$ defines a Hilbert space isomorphism from
the subspace of $L^2[-\pi,\pi]$ of odd functions onto $L^2[0,\pi]$
that carries the Hamiltonian of the planar pendulum to $H_2$.
Hence, we arrive again at the square potential with cosine
potential inside. By construction, the resulting isomorphism
$\Hi\equiv L^2(T,v\mr d t)^W \to L^2[0,\pi]$ coincides with
$\Gamma$.

\bre

The relation between our system and the quantum planar pendulum is
the quantum counterpart of the observation made above that the
reduced classical phase space of our system is isomorphic, as a
stratified symplectic space, to that of a spherical pendulum,
constrained to move with zero angular momentum, reduced relative
to rotations about the vertical axis. This system is manifestly
equivalent to that of  a planar pendulum reduced relative to
reflection about the vertical axis.

\ere

Now we discuss briefly the extension problem which arises in this context. 
Naive quantization after reduction on $\ctg\group$ 
fails because of the presence of singularities on $\pha$ .

The part of $\ctg \group$ to which
regular cotangent bundle reduction applies
is the cotangent bundle of the unreduced principal stratum
$\group\setminus\{\pm\II\}$. 
On this part, symplectic reduction leads to the
cotangent bundle of the quotient manifold, i.e., of the principal stratum
$\cfg_\prin$ but, beware, $\ctg\cfg_\prin$ 
is a proper subset of  the principal stratum
$\pha_\prin$ of the reduced phase space $\pha$ 
rather than being the entire stratum.
In the
parameterization of $\cfg$ chosen above, $\cfg_\prin$ corresponds to the open
interval $]0,\pi[$. Since the parameterization is an isometry when scaled via
$\scale$, canonical quantization of the kinetic energy then yields the
symmetric operator
 \beq\label{GnaiveDelta}
\scale^2~\frac{\mr d^2}{\mr d x^2}\,
 \eeq
on the Hilbert space $L^2[0,\pi]$ having as domain  the compactly
supported smooth functions on the open interval $]0,\pi[$. 
This leads to a naive quantization procedure away from the singularities
of $\cfg$. 

To arrive at a
well-defined quantum theory of the entire system including the singular subset
$\cfg_\sing$, one faces the problem of determining the self-adjoint
extensions of the operator \eqref{GnaiveDelta}, 
each of which defines a different quantum theory,
and to isolate one of these extensions as the
\lq correct\rq\ one. 
Thus, among the different
extensions, one has to pick 
one according to the
boundary conditions imposed on the wave functions
and the physical interpretation of the theory
will depend on  the choice of boundary conditions.
This is the problem studied in \cite{emmroeme}
in the situation where the classical configuration space is a cone over a
Riemannian manifold; see also \cite{DeserJackiw} and \cite{KayStuder} where
related questions are discussed under a more general perspective. When the
classical configuration space arises by reduction, the extension problem 
does not really arise, though, since by
reduction after quantization 
the kinetic energy operator is uniquely determined.
This was already observed in  \cite{Wren:Rieffel} in the context of
quantization by Rieffel induction. Indeed, in our situation, up to the shift by
$\scale^2$ which, in the case of \eqref{GnaiveDelta},
 can be obtained by the metaplectic
correction, $\Delta_2$ is a self-adjoint extension of \eqref{GnaiveDelta}.
According to Proposition \rref{Pcore}, this is the Friedrichs extension.

To conclude we speculate that some deeper insight into quantization after
reduction will, perhaps,  make the kinetic energy
operator unique in general as well.


\subsection{Eigenvalues and eigenstates}


For $\inco = 0$, i.\ e., in the strong coupling limit, in view
of \eqref{EWL} and \eqref{GEWCn}, the energy eigenvalues are given
by
$$
E_{n,\inco=0}
 =
\frac{\hbar^2}{2} \ve_n
 =
\frac{\hbar^2\scale^2}{2} n(n+2)
$$
and the corresponding normalized eigenfunctions are given by the
characters $\chi_n$. To solve the eigenvalue problem for
nonvanishing $\inco$ we carry $H$ via $\Gamma$ to $H_2$. Let
$$
\tinco = \frac{\inco}{\hbar^2\scale^2} \equiv \frac{1}{\hbar^2\scale^2 \coco^2}\,.
$$
According to \eqref{GH12} and \eqref{GDelta12}, on the core $D_2$ of $H_2$,
$$
H_2
 =
-\, \frac{\hbar^2\scale^2}{2}
 \left(
\frac{\mr d^2}{\mr d x^2}
 +
2 \tinco \cos(x) + \left(1 - 3\tinco\right)\right)\,,
$$
and so the stationary Schr\"odinger equation for $H_2$ reads
 \beq\label{GSreqf}
 \left(
\frac{\mr d^2}{\mr d x^2}
 +
2 \tinco \cos(x) + \left(\frac{2E}{\hbar^2\scale^2} + 1 -
3\tinco\right)
 \right)
\psi(x)
 =
0,
 \eeq
$E$ being the desired eigenvalue and $\psi\in D_2$ the
corresponding eigenfunction. The change of variable $y= (x - \pi)/2$
leads to the Mathieu equation
 \beq\label{GMathieu}
\frac{\mr d^2}{\mr d y^2} f(y) + (a-2q\cos(2y)) f(y) = 0\,,
 \eeq
where
 \beq\label{Gaq}
a = \frac{8 E}{\hbar^2\scale^2} + 4 - 12\tinco
 \,,~~~~~~
q = 4\tinco
 \,,
 \eeq
and $f$ is a Whitney smooth function on the interval $[-\pi/2,0]$
satisfying the boundary conditions
 \beq\label{Gboundcond}
f(-\pi/2) = f(0) = 0\,.
 \eeq
For the theory of the Mathieu equation and its solutions, called
Mathieu functions, see
\cite{Arscott,McLachlan,MeixnerSchaefke}. All we need is this: for
certain characteristic values of the parameter $a$, depending
analytically on $q$ and usually being denoted by $b_{2n+2}(q)$,
$n=0,1,2,\dots$, solutions satisying \eqref{Gboundcond} exist.
Given $a = b_{2n+2}(q)$, the corresponding solution is unique up
to a complex factor and can be chosen to be real-valued. It is
usually denoted by $\se_{2n+2}(y;q)$, where \lq$\se$\rq\  stands for
sine elliptic. For given $\tinco\geq 0$, let vectors
$\xi_n\in\Hi$ be defined by
 \beq\label{Gdefxi}
(\Gamma\xi_n)(x)
 =
(-1)^{n+1}\sqrt 2
 \left(
\se_{2n+2}\left(\frac{x - \pi}{2};4\tinco\right)
 \right)
 \,,~~~~~~~
n=0,1,2,\dots\,.
 \eeq
Since $\se_{2n+2}(y;0) = \sin((2n+2)y)$ the factor $(-1)^{n+1}$
ensures that for $\tinco=0$ we get $\xi_n = \chi_n$ exactly, and
not only up to a sign.

\btm\label{Teval}

For any $\tinco\geq 0$, the vectors $\xi_n\in\Hi$,
$n=0,1,2,\dots$, form an orthonormal basis of eigenvectors of $H$.
The corresponding eigenvalues are non-degenerate. They are given
by
$$
E_n
 =
\frac{\hbar^2\scale^2}{2}
 \left(
\frac{b_{2n+2}(4\tinco)}{4} + 3 \tinco - 1
 \right)\,.
$$

\etm

\bpf

This follows at once from the fact that, for any value of the
parameter $q$, the functions $\sqrt 2\,\se_{2n+2}(y;q)$, $n =
0,1,2,,\dots$, form an orthonormal basis in $L^2[-\pi/2,0]$, see
\cite[\S 20.5]{AbramowitzStegun}. Moreover, the characteristic
values satisfy $b_2(q)< b_4(q) < b_6(q) < \cdots$, see \cite[\S
20.2]{AbramowitzStegun}. Hence, for any value of $\tinco$ we have
$E_0 < E_1 < E_2 < \cdots$. \epf

\begin{figure}

\hspace*{-1.5cm}\epsfig{file=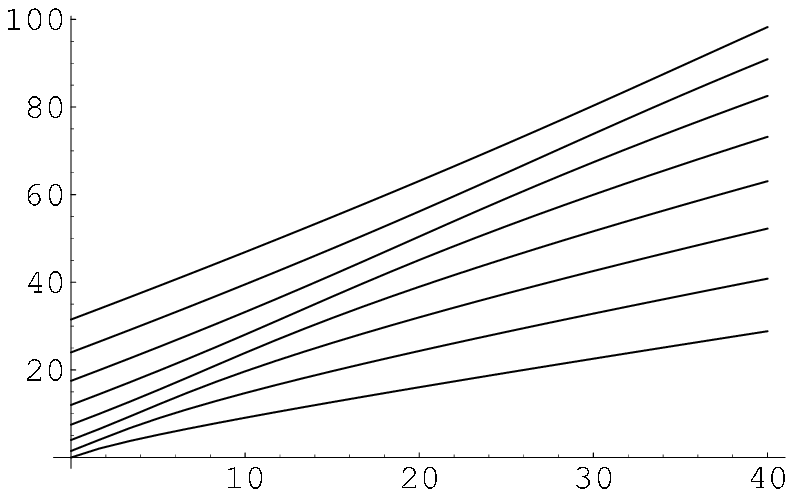,width=8cm}
\hspace*{-0.5cm}\epsfig{file=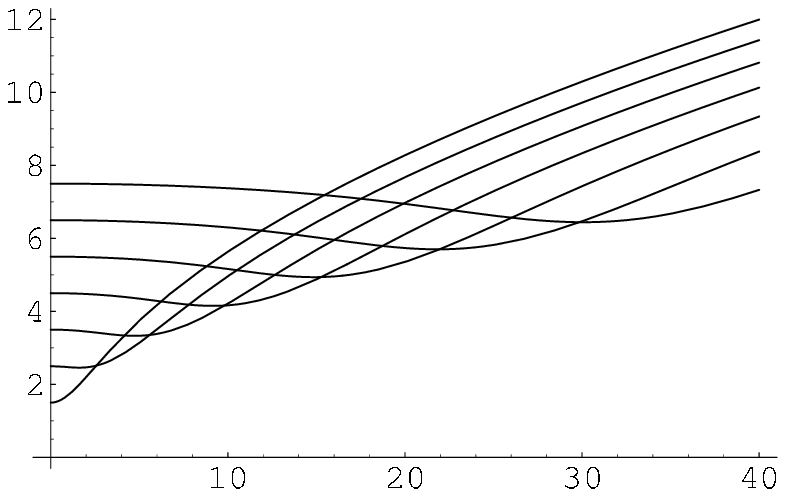,width=8cm}

\caption{\label{AbbEn} Energy eigenvalues (left) $E_n$ and
transition energy values $E_{n+1}-E_n$ (right) for $n=0,\dots,7$ in units of
$\hbar^2\scale^2$ as functions of $\tinco$.}

 \end{figure}

\begin{figure}
 \begin{center}

\unitlength1cm

 \begin{picture}(12,2.5)
 \put(0.1,-0.2){
 \marke{0,0}{cc}{\xi_0}
 \marke{4,0}{cc}{\xi_1}
 \marke{8,0}{cc}{\xi_2}
 \marke{12,0}{cc}{\xi_3}
 }
 \put(-2.8,0.2){
 \put(0,0){\epsfig{file=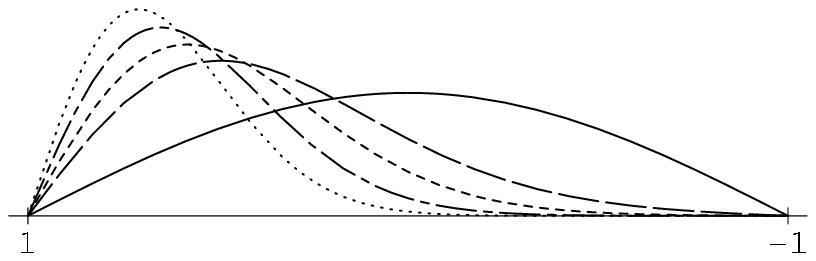,width=4.5cm}}
 \put(4,0){\epsfig{file=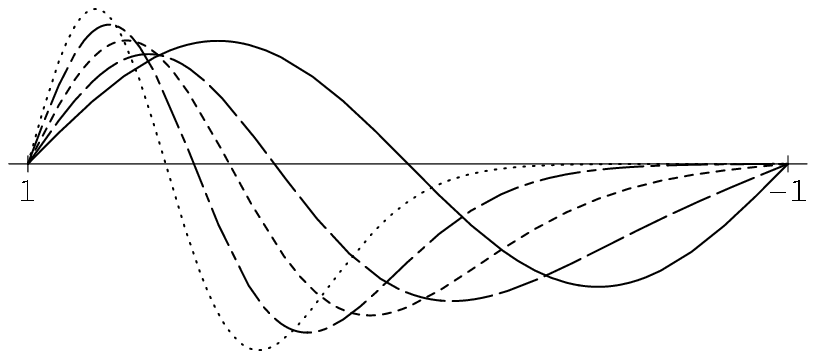,width=4.5cm}}
 \put(8,0){\epsfig{file=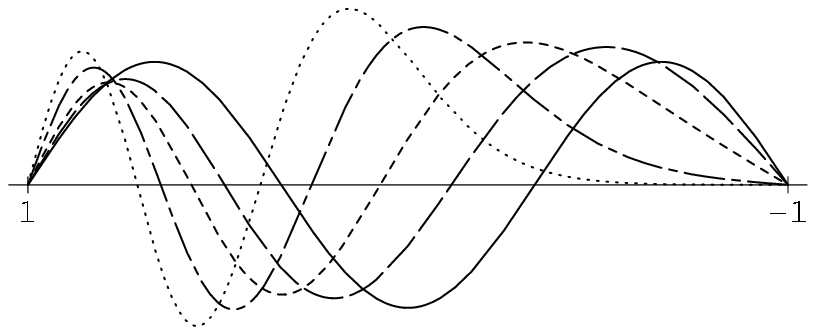,width=4.5cm}}
 \put(12,0){\epsfig{file=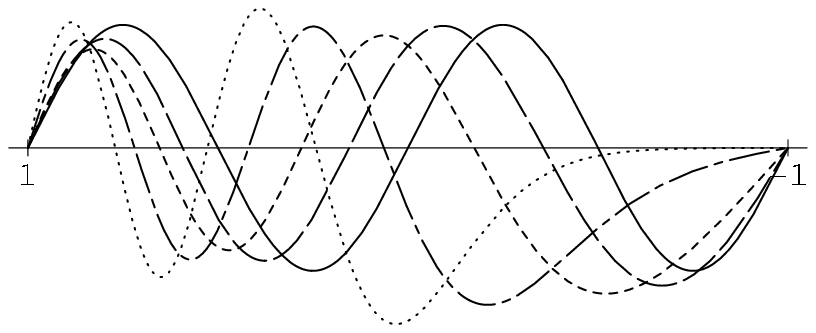,width=4.5cm}}
  }
 \end{picture}

 \end{center}

\caption{\label{Abbxin} Images of the energy eigenfunctions $\xi_0,\dots,\xi_3$,
under $\Gamma$, for $\tinco = 0$ (continuous line), $3$
(long dash), $6$ (short dash), $12$ (alternating short-long dash),
$24$ (dotted line).}

 \end{figure}

\noindent Figure \rref{AbbEn} shows the energy eigenvalues $E_n$
and the level separation $E_{n+1}-E_n$ for $n=0,\dots,8$ as
functions of $\tinco$. The transition energy values manifestly
reverse their order as $\inco$ increases. Figure \rref{Abbxin}
displays the images of eigenfunctions $\xi_n$, $n=0,\dots,3$, under
$\Gamma$ (i.e., the rescaled and shifted Mathieu functions
themselves), for $\tinco = 0,3,6,12,24$. The plots have been
generated by means of the built-in Mathematica functions {\tt
MathieuS} and {\tt MathieuCharacteristicB}.

\bre

The Schr\"odinger equation for the planar pendulum is solved in an
analogous way \cite{Condon}. From the discussion in Subsection
\rref{SSHam} it follows that the only difference is that in the
case of the pendulum, the function $f$ in \eqref{GMathieu} can be
any $\pi$-periodic smooth function on $\RR$. Then, in addition to
the family of $\pi$-periodic odd solutions given by the functions
$\se_{2n+2}(y;q)$ and their characteristic values $b_{2n+2}(q)$
there is a family of $\pi$-periodic even solutions which are
usually denoted by $\ce_{2n+2}$ (for \lq cosine elliptic\rq). The
corresponding characteristic values are usually denoted by
$a_{2n+2}(q)$. For any value of $q$, $a_2(q) < b_2(q) < a_4(q) <
b_4(q) <\cdots$. Thus, precisely every second eigenstate of the
planar pendulum emerges in our system. In particular, the ground
state of our system does not correspond to the ground state but to
the first excited state of the planar pendulum.

\ere

\bre

According to Remark \rref{Remmodel}, Theorem \rref{Teval} yields
the solutions to the stationary Schr\"odinger equation for quantum
Yang-Mills theory on $S^1$ when the self-interaction is described
by the potential in \eqref{GHaFn}. In particular, in this simple
model we have constructed the vacuum and all excited states, for
arbitrary values of the coupling constant.

\ere


\section{Expectation values of the costratification orthoprojectors for
$\SU(2)$}

\label{Sexpect}


The most elementary observables associated with the
costratification are the orthoprojectors $\Pi_i$ onto,
respectively, the subspaces $\Hi_i$, $i=\pm,\sing$. The
expectation value of $\Pi_i$ in a state $\psi$ yields the
probability that the system prepared in this state is measured in
the subspace $\Hi_i$. We determine the expectation values of
$\Pi_i$ in the energy eigenstates, i.~e.,
$$
P_{i,n} := \langle \xi_n | \Pi_i \xi_n \rangle = \| \Pi_i\xi_n
\|^2
 \,,~~~~~~
i=\sing,\pm\,.
$$
Then, we derive approximations for these expectation values for
strong and weak coupling.


\subsection{Expectation values} \label{SSexpect}

\begin{figure}

\unitlength1cm

 \begin{picture}(12,11.5)
 \put(-1,8){
 \put(0,0){\epsfig{file=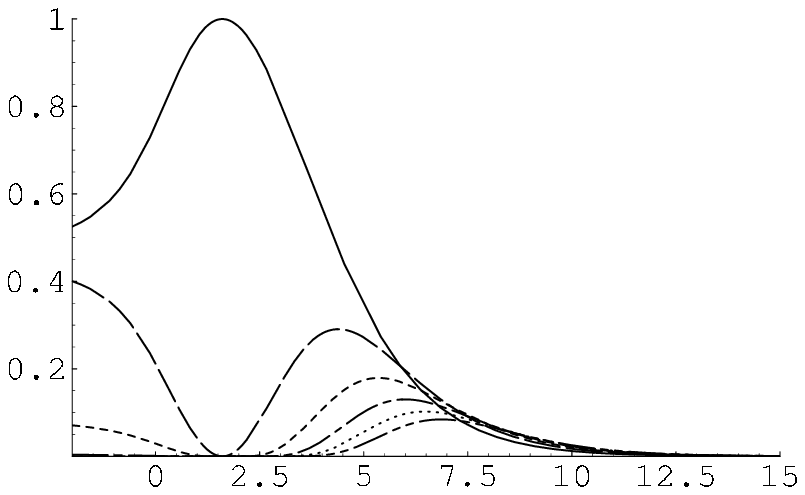,width=6cm,height=2cm}}
 \put(5,0){\epsfig{file=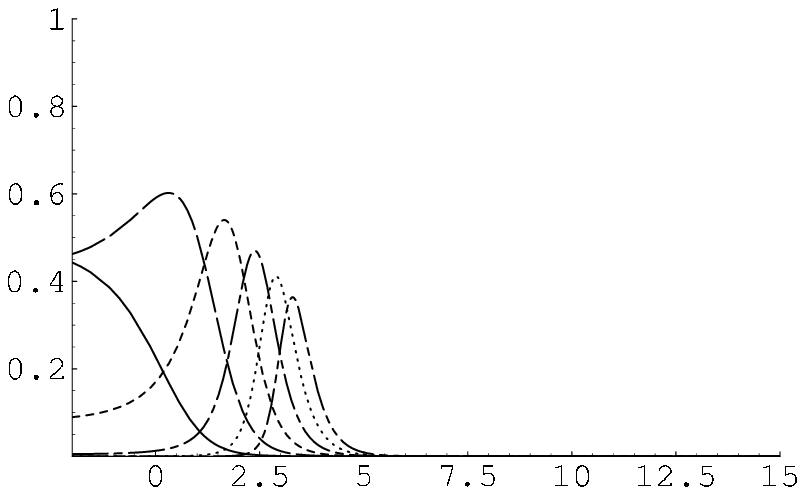,width=6cm,height=2cm}}
 \put(10,0){\epsfig{file=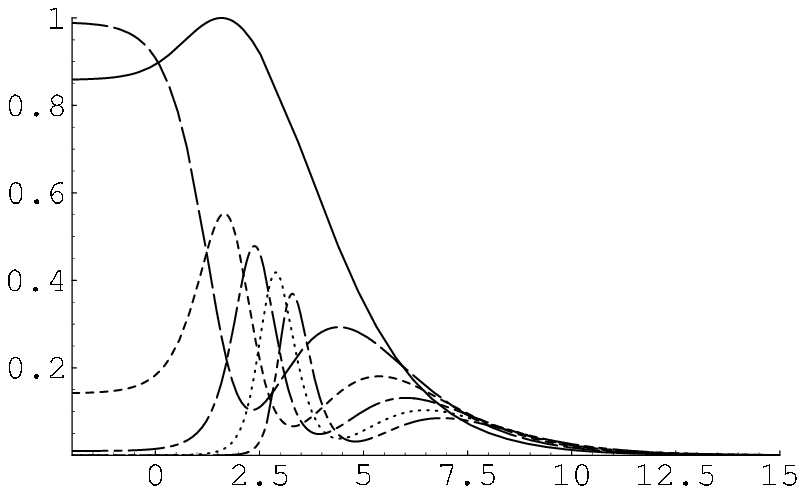,width=6cm,height=2cm}}
  }
 \put(-1,4.5){
 \put(0,0){\epsfig{file=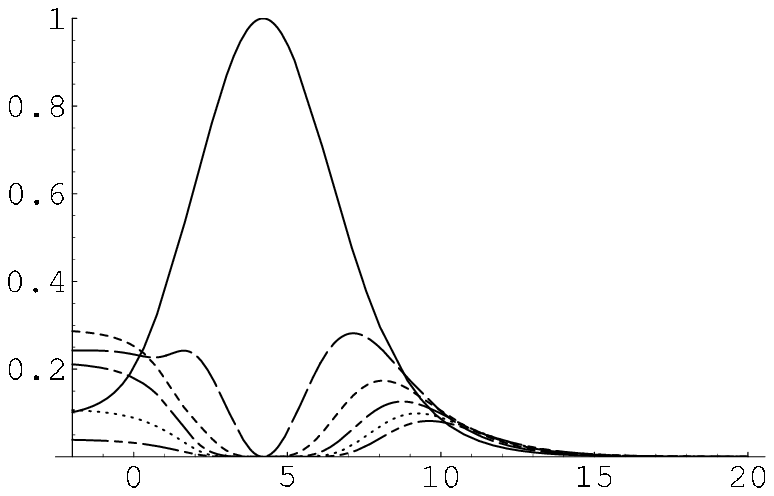,width=6cm,height=2cm}}
 \put(5,0){\epsfig{file=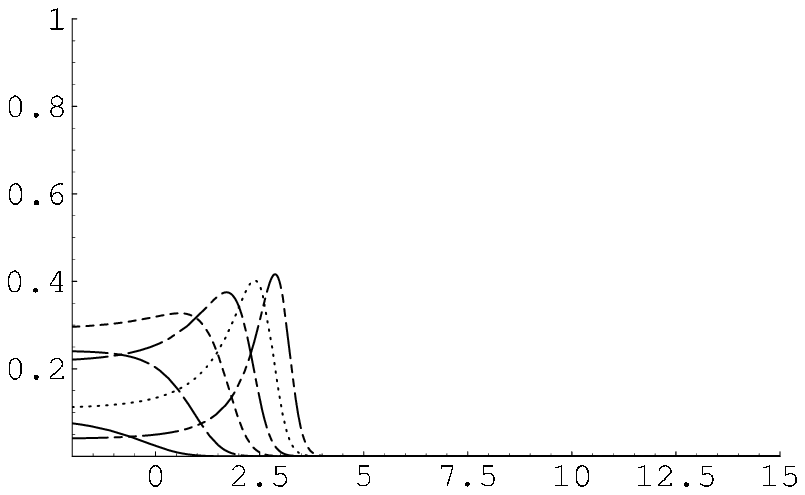,width=6cm,height=2cm}}
 \put(10,0){\epsfig{file=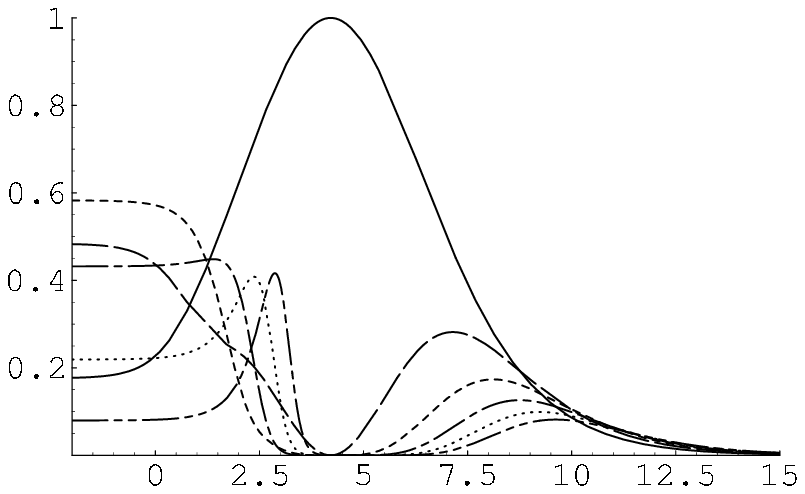,width=6cm,height=2cm}}
  }
 \put(-1,1){
 \put(0,0){\epsfig{file=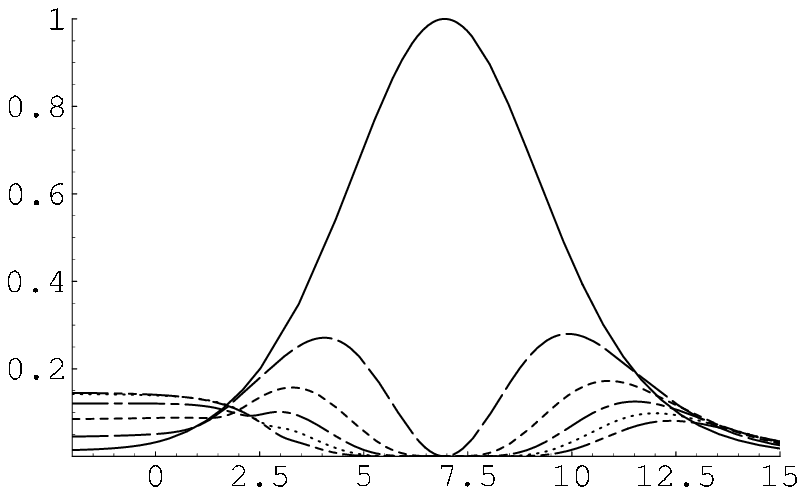,width=6cm,height=2cm}}
 \put(5,0){\epsfig{file=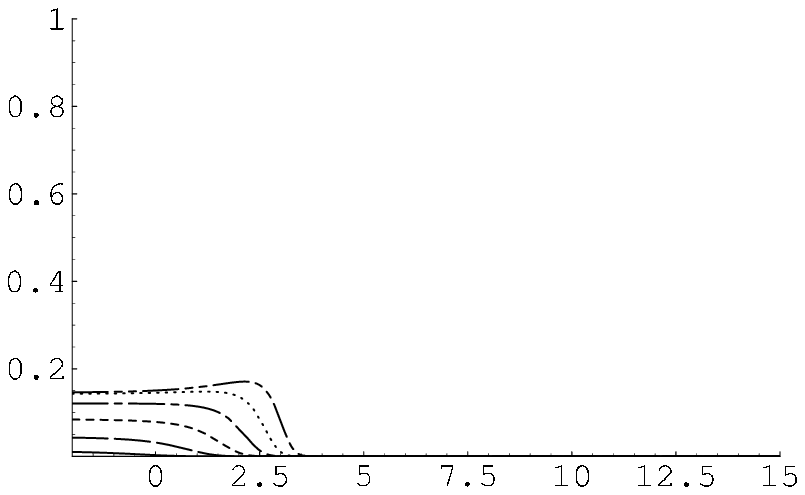,width=6cm,height=2cm}}
 \put(10,0){\epsfig{file=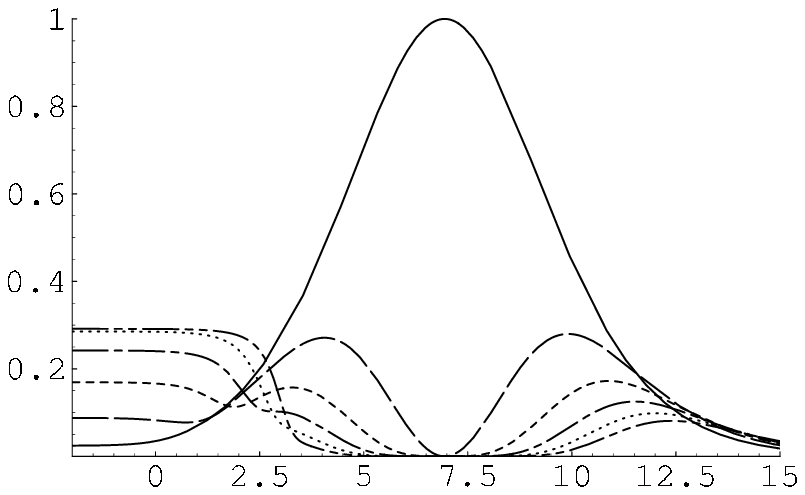,width=6cm,height=2cm}}
  }
 \put(1,7.5){
 \marke{0,0}{cl}{P_{+,n}\,,~~\hbar\scale^2 = \frac 1 2}
 \marke{5,0}{cl}{P_{-,n}\,,~~\hbar\scale^2 = \frac 1 2}
 \marke{10,0}{cl}{P_{\sing,n}\,,~~\hbar\scale^2 = \frac 1 2}
 }
 \put(1,4){
 \marke{0,0}{cl}{P_{+,n}\,,~~\hbar\scale^2 = \frac 1 8}
 \marke{5,0}{cl}{P_{-,n}\,,~~\hbar\scale^2 = \frac 1 8}
 \marke{10,0}{cl}{P_{\sing,n}\,,~~\hbar\scale^2 = \frac 1 8}
 }
 \put(1,0.5){
 \marke{0,0}{cl}{P_{+,n}\,,~~\hbar\scale^2 = \frac 1 {32}}
 \marke{5,0}{cl}{P_{-,n}\,,~~\hbar\scale^2 = \frac 1 {32}}
 \marke{10,0}{cl}{P_{\sing,n}\,,~~\hbar\scale^2 = \frac 1 {32}}
 }
 \end{picture}

 \caption{\label{AbbPin} Expectation values $P_{+,n}$, $P_{-,n}$
and $P_{\sing,n}$ for $n=0$ (continuous line), $1$ (long dash), $2$
(short dash), $3$ (long-short dash), $4$ (dotted line) and $5$
(long-short-short dash), plotted over $\log\tinco$ 
for $\hbar\scale^2 = \frac 1 2, \frac 1 8, \frac 1 {32}$.}

 \end{figure}

According to \eqref{Goproj},
 \beq\label{GPin}
P_{\pm,n}
  =
|\langle \xi_n | \psi_\pm \rangle|^2
 \,,~~~~~~
P_{\sing,n}
  =
|\langle \xi_n | \psi_\even \rangle|^2
 +
|\langle \xi_n | \psi_\odd \rangle|^2 .
 \eeq
As $\se_{2n+2}$ is odd and $\pi$-periodic, it can be expanded as
$$
\se_{2n+2}(y;q) = \sum_{k=0}\nolimits^\infty B^{2n+2}_{2k+2}(q)
\sin((2k+2)y)\,,
$$
where $B^{2n+2}_{2k+2}(q)$ refers to the Fourier coefficients. The
Fourier coefficients satisfy certain recurrence relations, see
\cite[\S20.2]{AbramowitzStegun}. Using \eqref{GGammachi}, we find
 \beq\label{Gscaproxichi}
\langle \xi_n | k \rangle = (-1)^{n+k}
B^{2n+2}_{2k+2}(4\tinco)\,.
 \eeq
Then \eqref{Gpsip}--\eqref{Gpsising} yield
 \beqa\label{Gxipsip}
\langle \xi_n | \psi_+ \rangle
 & = &
\frac{(-1)^n}{N}
 \sum\nolimits_{k=0}^\infty
(-1)^k\,(k+1)\,\mr e^{-\hbar\scale^2(k+1)^2/2} \,
B^{2n+2}_{2k+2}(4\tinco) ,
\\ \label{Gxipsim}
\langle \xi_n | \psi_- \rangle
 & = &
\frac{(-1)^n}{N}
 \sum\nolimits_{k=0}^\infty
(k+1)\,\mr e^{-\hbar\scale^2(k+1)^2/2} \, B^{2n+2}_{2k+2}(4\tinco) ,
\\ \label{Gxipsi0}
\langle \xi_n | \psi_\even \rangle
 & = &
\frac{(-1)^n}{N_\even}
 \sum\nolimits_{k=0}^\infty
(2k+1) \mr e^{-\hbar\scale^2(2k+1)^2/2} B^{2n+2}_{4k+2}(4\tinco)
 \,,
\\ \label{Gxipsi1}
\langle \xi_n | \psi_\odd \rangle
 & = &
-\,\frac{(-1)^n}{N_\odd}
 \sum\nolimits_{k=0}^\infty
(2k+2) \mr e^{-\hbar\scale^2(2k+2)^2/2}
B^{2n+2}_{4k+4}(4\tinco)\,.
 \eeqa
Together with \eqref{GPin}, this yields formulas for the $P_{i,n}$'s,
$i=\sing,\pm$.
We do not spell them out, since they do not lead to
significant simplification.
The functions $P_{i,n}$ depend on the parameters $\hbar$,
$\scale^2$ and $\inco$ only via the combinations $\hbar\scale^2$ and
$\tinco = \inco/(\hbar^2\scale^2)$.
Figure \rref{AbbPin} displays
the $P_{i,n}$, $n=0,\dots,5$, as functions of $\tinco$ for three specific
values of $\hbar\scale^2$, thus treating $\tinco$ and $\hbar\scale^2$
as independent parameters. This is appropriate for the discussion of
the dependence of $P_{i,n}$ on the coupling parameter $g$ for fixed
values of $\hbar$ and $\scale^2$. The plots have been generated by
Mathematica through numerical integration.

Perhaps the most impressive feature is the dominant peak of
$P_{+,0}$ which is enclosed by less prominent maxima of the other
$P_{+,n}$ and moves to higher $\tinco$ when $\hbar\scale^2$
decreases. In other words, for a certain value of the coupling constant, the
state $\psi_+$ which spans $\Hi_+$ seems to coincide almost
perfectly with the ground state. If the two states coincided completely
then \eqref{Gscaproxichi} would imply that, for a certain value of $q$,
the coefficients $B^{2n+2}_{2k+2}(q)$ would be given by
$(-1)^{n+k}\frac 1 N (k+1)\mr e^{-\hbar\scale^2(k+1)^2/2}$.
However, this is not true; the latter expressions do not
satisfy the recurrence relations valid for the coefficients
$B^{2n+2}_{2k+2}(q)$ for any value of $q$. Another interesting phenomenon is
that, for decreasing $\hbar\scale^2$, the maxima of $P_{-,n}$ move
to lower $\tinco$ and the subsequent descent becomes steeper.

Next, we will derive approximations for the $P_{i,n}$'s for small
and large values of $\tinco$. When $\hbar$ and $\scale$ are fixed,
this corresponds to strong and weak coupling, as appropriate. The
strong coupling approximation will provide a resolution of the
first crossings of the graphs of the $P_{i,n}$. The weak coupling
approximation will allow us to analyze the position and the height
of the dominant peak of $P_{+,0}$ as well as of the subsequent
maxima of the other $P_{+,n}$'s. A detailed study of the maxima of
the $P_{-,n}$'s and of the behaviour of the $P_{+,n}$'s in the intermediate
region between strong and weak coupling remains as a future task.


\subsection{Strong coupling approximation}

\begin{figure}

\epsfig{file=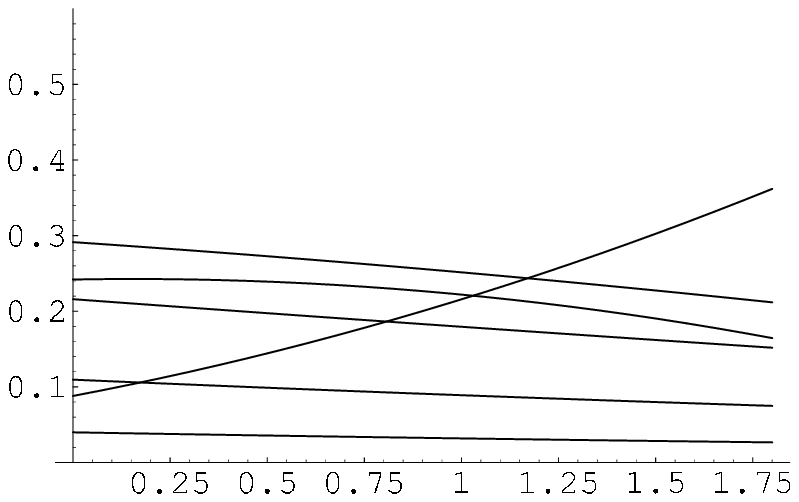,width=5cm}
\epsfig{file=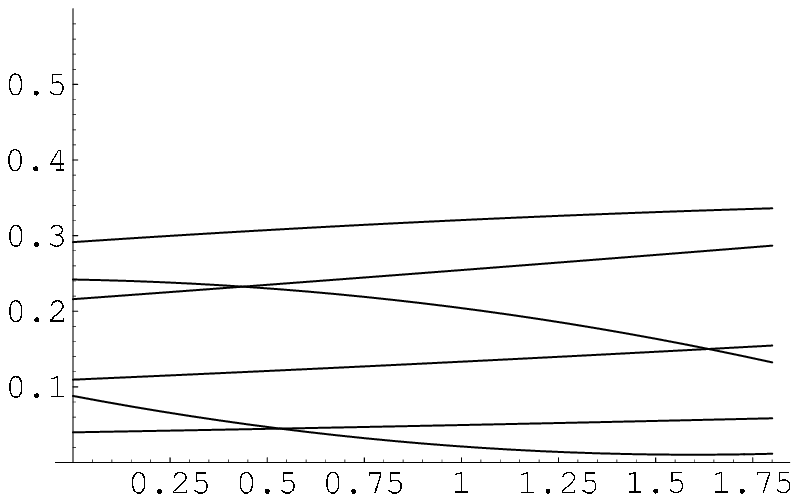,width=5cm}
\epsfig{file=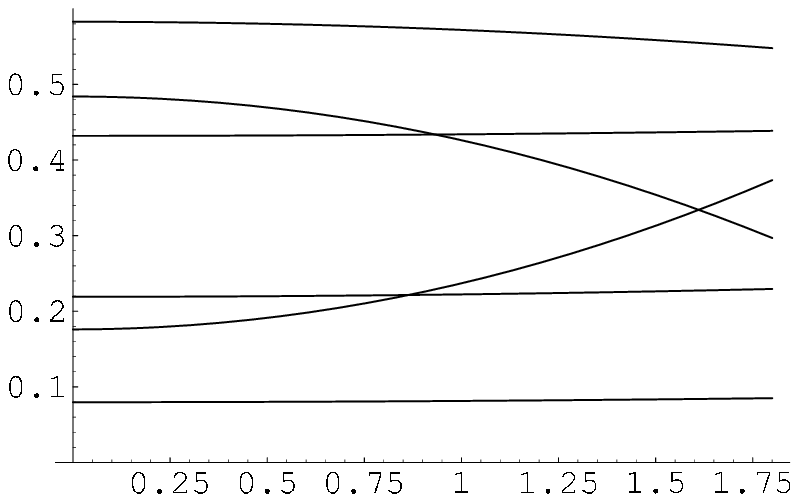,width=5cm}

\caption{\label{AbbPinstrong} Quadratic approximations for
$P_{+,n}$, $P_{-,n}$, $P_{0,n}$ (from left to right),
$n=0,\dots,5$, $\hbar\scale^2=\frac 1 8$, plotted over $\tinco$.}

 \end{figure}

In the region of strong coupling, i.~e., for large $g$, $\tinco$ is
small, at least when the parameter $\hbar\scale^2$ is fixed. Power
series expansions for the characteristic values $b_{2n+2}(q)$ in
$q$ about $q = 4\tinco = 0$ can be found, e.~g., in \cite[\S
20.2.25]{AbramowitzStegun}. They immediately provide expansions
for the energy eigenvalues. We do not spell out the latter here,
because we are merely interested  in approximations of the
expectation values $P_{i,n}$, $i=\pm,\sing$. Quadratic
approximations for the Fourier coefficients $B^{2n+2}_{2k+2}(q)$
in $q$ can be read off from \cite[\S 2.25, (42)]{MeixnerSchaefke}:
For the central coefficients, this yields
$$
 \begin{array}{rclcrcl}
B^2_2(4\tinco) & = & 1 - \frac{1}{18} \, \tinco^2 + O(\tinco^3)
 \,, &  &
B^{2n+2}_{2n+2}(4\tinco)
 & = &
1 - \frac{(2n+2)^2 + 1}{2((2n+2)^2-1)^2} \, \tinco^2 + O(\tinco^3)
  \,,~~ n \geq 1\,.
 \end{array}
$$
For the next-to-central coefficients,
$$
 \begin{array}{rclcrcl}
B^{2n+2}_{2n}(4\tinco) & = & \frac{1}{(2n+1)}\, \tinco +
O(\tinco^3)
 \,, & \hspace*{-0.5cm} &
B^{2n+2}_{2n+4}(4\tinco) & = & - \frac{1}{(2n+3)} \tinco +
O(\tinco^3)
 \,,
\\
B^{2n+2}_{2n-2}(4\tinco) & = & \frac{1}{4n(2n+1)} \tinco^2 +
O(\tinco^3)
 \,, & &
B^{2n+2}_{2n+6}(4\tinco) & = & \frac{1}{2(2n+3)(2n+4)} \tinco^2 +
O(\tinco^3)
 \,,~~ n\geq 0\,.
 \end{array}
$$
All the other coefficients are of order $O(\tinco^3)$. Using
\eqref{GPin} and \eqref{Gxipsip}--\eqref{Gxipsi1} we obtain
$$
P_{\pm,n}
 =
\frac{A_{n,0}^2}{N^2}
 \pm
\frac{2 A_{n,0} A_{n,1}}{N^2}~ \tinco
 +
\frac{A_{n,1}^2 + A_{n,0} A_{n,2}}{N^2}~ \tinco^2
 +
O(\tinco^3)
$$
and, for $n$ even, we get
$$
P_{\sing,n} = \frac{A_{n,0}^2}{N_\even^2}
 + \left(
\frac{A_{n,1}^2}{N_\odd^2} + \frac{A_{n,0}A_{n,2}}{N_\even^2}
 \right) \tinco^2 + O(\tinco^3)
 \,,
$$
whereas, for $n$ odd, in this expression,
one has to interchange $N_\even$ and $N_\odd$. The coefficients are
$$
 \begin{array}{rcl}
A_{n,0} & = & (n+1) \mr e^{-\hbar\scale^2(n+1)^2/2}
 \,,~~~~~~
n\geq 0\,,
\\
A_{n,1}
 & = &
\frac{(n+2)\mr e^{-\hbar\scale^2(n+2)^2/2}}{2n+3}
 -
\frac{n\mr e^{-\hbar\scale^2 n^2/2}}{2n+1}
 \,,~~~~~~
n\geq 0\,,
\\
A_{0,2}
 & = &
\frac{\mr e^{-9\hbar\scale^2/2}}{4} - \frac{\mr
e^{-\hbar\scale^2/2}}{9}\,,
\\
A_{n,2}
 & = &
\frac{(n-1)\mr e^{-\hbar\scale^2(n-1)^2/2}}{2n (2n+1)}
 +
\frac{(n+3)\mr e^{-\hbar\scale^2(n+3)^2/2}}{(2n+3)(2n+4)}
\\
 & &
\phantom{\frac{(n-1)\mr e^{-\hbar\scale^2(n-1)^2/2}}{2n (2n+1)}}
 -
(n+1) \mr e^{-\hbar\scale^2(n+1)^2/2}
 \left(
\frac{1}{(2n+1)^2} + \frac{1}{(2n+3)^2}
 \right)
 \,,~~~~~~
n\geq 1\,.
 \end{array}
$$
For $\hbar\scale^2 = \frac 1 8$,
plots of the quadratic approximations of the $P_{i,n}$'s, $i=\pm,\sing$,
are shown in Figure
\rref{AbbPinstrong}, for $n=0,\dots,5$ and $\tinco$ ranging between
$0$ and $0.2$. Here the approximation has a relative error of less
than $0.01$. The plots yield, in particular, a resolution of the
first crossings of the graphs of the $P_{i,n}$'s in the bottom line
of Figure \rref{AbbPin}.

For very strong
coupling, the state $\xi_2$ rather than the ground state has
the highest probability to be measured in $\Hi_+$.
 In fact the ground state is excelled by all $\xi_n$ with $n\leq 4$.
(This follows of course directly from consideration of the case $\inco
= 0$, where $\xi_n = \chi_n$.) The precise order of the
expectation values in this region is
$$
P_{i,2} \geq P_{i,1} \geq P_{i,3} \geq P_{i,4} \geq P_{i,0} \geq
P_{i,5} \geq P_{i,6} \geq \cdots\,,~~~~~~ i=\sing,\pm\,.
$$
On the other hand, the probabilities $P_{i,0}$ of the ground state
change most rapidly as $\inco$ increases. In particular, $P_{+,0}$
has overtaken all other probabilities already for $\inco = 0.2$.


\subsection{Weak coupling approximation}


Similarly to the approximation of a classical planar pendulum by a
classical harmonic oscillator, for excitations that are small
compared with the length of the pendulum, the quantum planar
pendulum can be approximated by a quantum harmonic oscillator for
energy values that are small compared with the range of the potential
\cite{AldrovandiLealFerreira, BakerBlackburnSmith, Condon, PradhanKhare}. We
use this procedure to obtain approximations for the energy eigenfunctions
$\xi_n$ and, from these, approximations for the expectation values $P_{i,n}$,
$i=\pm,\sing$, for large $\tinco$ and small $n$.

Consider the Schr\"odinger equation for $H_2$ in \eqref{GSreqf}.
Making the change of variable $z = \sqrt[4]{\tinco}\,x$ we obtain the
equation
$$
\left(\frac{\mr d^2}{\mr d z^2}
 +
2\tinco \cos(z/\sqrt[4]{\tinco})
 +
 \left(
\frac{1}{\sqrt\tinco}
 \left(
\frac{2E}{\hbar^2\scale^2}
 +
1
 \right)
 -
3 \sqrt\tinco
 \right)\right)
f(z) = 0
$$
where $f$ may be a Whitney smooth function on the interval
$[0,\sqrt[4]\tinco\pi]$ satisfying the boundary conditions $f(0) =
f(\sqrt[4]\tinco\pi) = 0$. Replacing the cosine with its second
order Taylor expansion we obtain the Schr\"odinger equation
 \beq\label{GSreqhO}
 \left(
\frac{\mr d^2}{\mr d z^2}
 -
z^2
 +
2\epsilon
 \right) f(z) = 0
 \eeq
of the harmonic oscillator with unit frequency, where
\beq\label{frequency}
\epsilon
 =
\frac{1}{\sqrt\tinco}
 \left(
\frac{E}{\hbar^2\scale^2}
 +
\frac 1 2
 \right)
 -
\frac{\sqrt\tinco}{2}\,.
 \eeq
For large $\tinco$ and small energies, the solutions of either
equation are concentrated about $x = z/\sqrt[4]{\tinco} \sim 0$.
Under these circumstances, restriction to the interval
$[0,\sqrt[4]\tinco\pi]$ of solutions of \eqref{GSreqhO} satisfying
$f(0) = 0$ yields satisfactory approximations for solutions of
\eqref{GSreqf}. The appropriate solutions of \eqref{GSreqhO} are
well known to be
$$
f(z) = H_{2n+1}(z) \mr e^{-z^2/2}
 \,,~~~~~~
\epsilon
 =
2n + \frac 3 2
 \,,~~~~~~
n=0,1,2,\dots\,,
$$
where
$$
H_{2n+1}(z)
 =
\sum_{r=0}^n \frac{(-1)^{n+r} (2n+1)! (2 z)^{2r+1}}{(n-r)! (2r+1)!}
$$
are the odd degree Hermite polynomials. Define vectors
$\xi^{(\infty)}_n\in\Hi$ by
 \begin{eqnarray}\label{Gdefxiinf}
(\Gamma\xi^{(\infty)}_n)(x)
 & = &
(-1)^n\,N^{(\infty)}_n
 ~
H_{2n+1}\!\left(\sqrt[4]{\tinco}\, x\right)
 ~
\mr e^{-\sqrt{\tinco}\, x^2/2}
 \,,
\\ \label{GNnhO}
N^{(\infty)}_n
 & = &
\frac{\pi^{1/4}\,\tinco^{1/8}}{2^{n+1}\,\sqrt{(2n+1)!}}\,,
 \end{eqnarray}
where the choice of sign is dictated by that for the $\xi_n$'s, see
\eqref{Gdefxi}. The $\xi^{(\infty)}_n$'s form a
basis of $\Hi$. Substituting for $\epsilon$  the
right-hand side of \eqref{frequency}, we obtain the energy values
$$
E^{(\infty)}_n
 =
\frac{\hbar^2\scale^2}{2}
 \left(
\tinco + (4n+3) \sqrt\tinco -1\right)\,.
$$
The $E^{(\infty)}_n$'s and the $\xi^{(\infty)}_n$'s yield
approximations for the true energy eigenvalues $E_n$ and for the
eigenfunctions $\xi_n$ of our model for large $\tinco$ and small
$n$. Note that the $\xi_n$'s are neither orthogonal nor normalized, because the 
functions $\Gamma\xi^{(\infty)}_n$ are orthogonal over the interval $[0,\infty]$
rather than the interval $[0,\pi]$ and the normalization factor $N^{(\infty)}_n$
is therefore also chosen so that the functions are normalized
over the interval $[0,\infty]$. The deviation from being orthonormal is however
negligible for small $n$ and large $\tinco$. 

\begin{figure}

 \begin{center}

\unitlength1cm

 \begin{picture}(12,2.7)

 \put(0.1,-0.2){
 \marke{0,0}{cc}{n=0}
 \marke{4,0}{cc}{n=1}
 \marke{8,0}{cc}{n=2}
 \marke{12,0}{cc}{n=3}

 \marke{-1.05,2.45}{cb}{\scriptscriptstyle a}
 \marke{-0.525,2.45}{cb}{\scriptscriptstyle b}
 \marke{-0.05,2.45}{cb}{\scriptscriptstyle c}

 \marke{3.65,2.1}{cb}{\scriptscriptstyle a}
 \marke{4,2.1}{cb}{\scriptscriptstyle b}
 \marke{4.5,2.1}{cb}{\scriptscriptstyle c}

 \marke{7.7,1.72}{cb}{\scriptscriptstyle a}
 \marke{8.15,1.72}{cb}{\scriptscriptstyle b}
 \marke{8.65,1.72}{cb}{\scriptscriptstyle c}

 \marke{11.8,1.8}{cb}{\scriptscriptstyle a}
 \marke{12.3,1.8}{cb}{\scriptscriptstyle b}
 \marke{12.8,1.8}{cb}{\scriptscriptstyle c}
 }
 \put(-3,0.2){
 \put(0,0){\epsfig{file=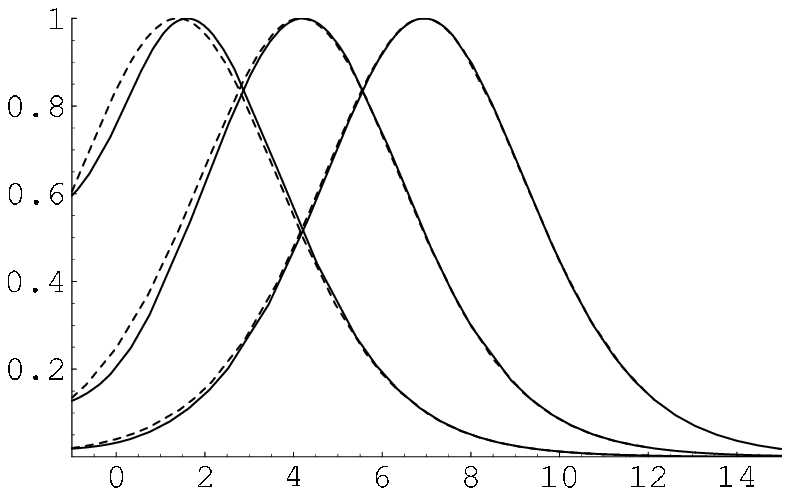,width=4.5cm}}
 \put(4,0){\epsfig{file=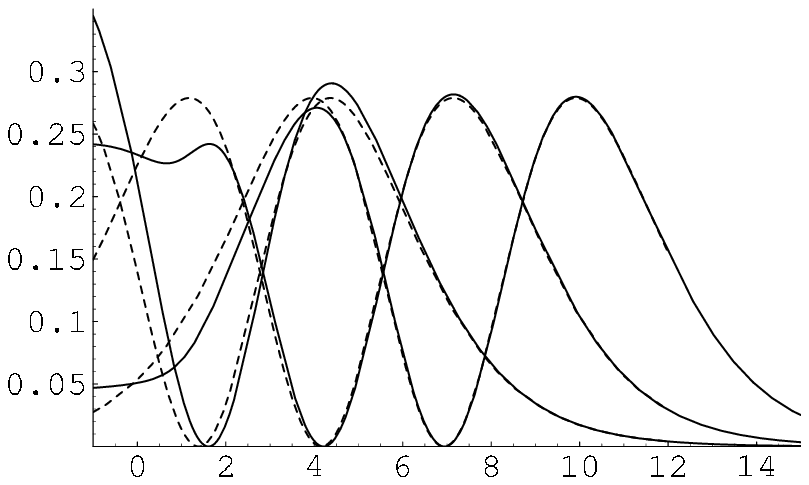,width=4.5cm}}
 \put(8,0){\epsfig{file=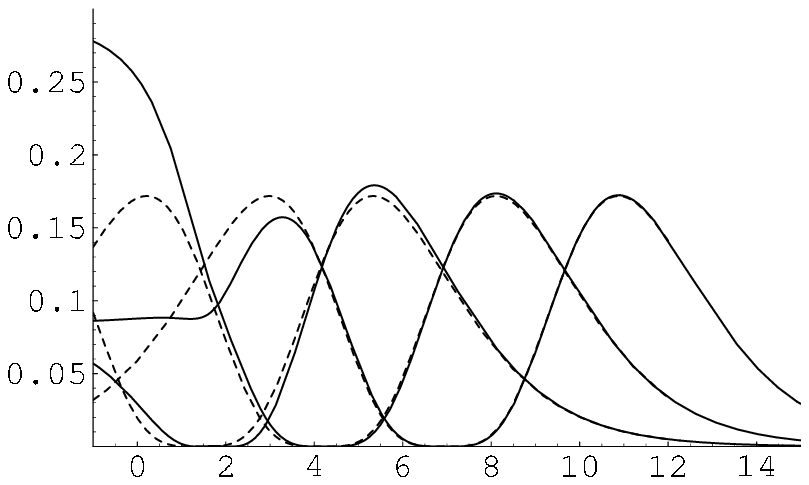,width=4.5cm}}
 \put(12,0){\epsfig{file=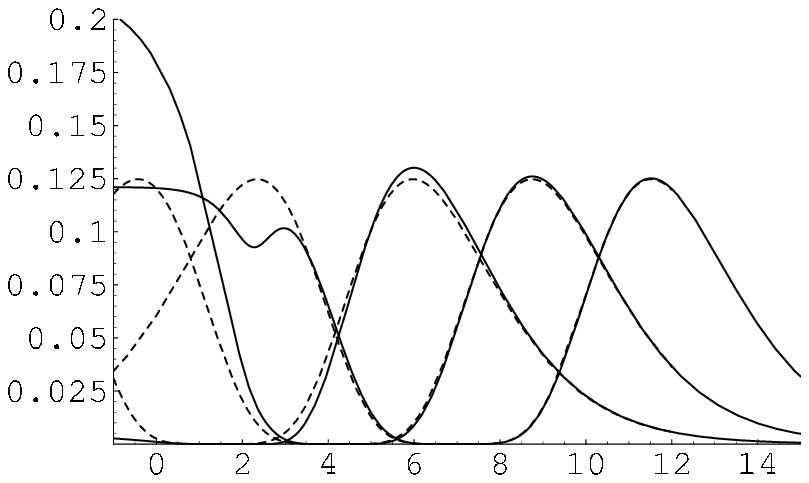,width=4.5cm}}
  }
 \end{picture}

 \end{center}

\caption{\label{AbbPprat}Exact values of $P_{+,n}$ (continuous
lines) and approximations $P_{+,n}^{(\infty)}$ (dashed lines) as
functions of $\tinco$ on a logarithmic scale for $\hbar\scale^2 =
\frac 1 2 (a), \frac 1 8 (b), \frac 1 {32} (c)$ and $n=0,\dots,3$.}

 \end{figure}

To compute the scalar products $\langle\chi_k,\xi^{(\infty)}_n\rangle$, we use
\eqref{GGammachi}  
and \eqref{Gdefxiinf} and move the upper bound of the resulting integral from
$\pi$ to infinity, which is again justified for large $\tinco$ and small $n$.
The result is
 \beq\label{Gscaproxichi2}
\langle\chi_k,\xi^{(\infty)}_n\rangle
 =
\frac{2^{-n} \, \pi^{-1/4}}{\sqrt{(2n+1)!}}
 \,
\tinco^{-1/8}
 \,
H_{2n+1}\big((k+1)\,\tinco^{-1/4}\big)
 \,
\mr e^{-(k+1)^2 \, \tinco^{-1/2}/2}
 \,.
 \eeq
This formula is also a consequence of \eqref{Gscaproxichi} and the
asymptotic expansion of the Fourier coefficients
$B^{2n+2}_{2k+2}(q)$ for large $q$ given in
\cite[\S 2.333]{MeixnerSchaefke}. Using \eqref{Gscaproxichi2} and
writing out the formula of the Hermite polynomials, we obtain
 \beq\label{GscaprohoHerm}
\langle \psi_i,\xi_n^{(\infty)}\rangle
 =
(-1)^n \frac{\sqrt{(2n+1)!}}{2^n \pi^{1/4} N_i}
 \sum_{r=0}^n
\frac{(-1)^r \, 2^{2r+1}\, \tinco^{-(4r+3)/8}}{(n-r)! (2r+1)!}
~~\Sigma_i^r\!\!\left(\frac{\hbar\scale^2 +
\tinco^{-1/2}}{2}\right)\,,
 \eeq
where $i=\pm,\even,\odd$, $N_\pm \equiv N$, and
 \begin{align*}
\Sigma_+^r(b) & = \sum_{k=1}^\infty k^{2r+2} \mr e^{-b k^2}
 \,, &
\Sigma_-^r(b) & = \sum_{k=1}^\infty (-1)^{k+1} k^{2r+2} \mr e^{-b
k^2}\,.
\\
\Sigma_\even^r(b) & = \sum_{k\text{ odd}} k^{2r+2} \mr e^{-bk^2}
 \,, &
\Sigma_\odd^r(b) & = \sum_{k\text{ even}} k^{2r+2} \mr e^{-b
k^2}\,.
 \end{align*}
Expressing the sums in terms of the theta constant $\theta_3$, see
\eqref{Gdeftheta}, we obtain
 \begin{align*}
\Sigma_+^r(b)
 & =
\frac 1 2 \frac{\mr d^{r+1}}{\mr d (-b)^{r+1}} \theta_3(\mr
e^{-b})
 \,, &
\Sigma_-^r(b) & = \Sigma_+^r(b) - 2^{2r+3}  \Sigma_+^r(4b)
 \,,
\\
\Sigma_\even^r(b) & = \Sigma_+^r(b) - 4^{r+1} \Sigma_+^r(4b)
 \,, &
\Sigma_\odd^r(b) & = 4^{r+1} \Sigma_+^r(4b)
 \,.
 \end{align*}
Substituting in \eqref{GscaprohoHerm}, for $\Sigma_i^r$, the right-hand side of
each of these identities as appropriate and taking the square, we
arrive at the harmonic oscillator approximations of $P_{\pm,n}$ and
$P_{0,n}$. These approximations are hard to handle, however, as they contain
higher derivatives of the theta constant w.r.t.\ the nome. Instead, we
use the approximation
 \beq\label{Gapproxtheta}
\theta_3(e^{-y}) = \sqrt\pi y^{-1/2} + \dots\,,
 \eeq
valid for small $y$ and hence for small $\hbar\scale^2$ and
large $\tinco$. Even for $y=1$, this approximation has a relative error of only
$10^{-4}$. In this approximation,
 \beq\label{GSigma+r}
\Sigma_+^r(b)
 =
\sqrt\pi\frac{(2r+1)!}{4^{r+1} r!} b^{-(2r+3)/2}
 \,,~~~~~~
\Sigma_-^r(b) = 0
 \,,~~~~~~
\Sigma_\even^r(b) = \Sigma_\odd^r(b) = \frac 1 2 \Sigma_+^r(b)\,,
 \eeq
and
$$
N
 =
\sqrt{\Sigma_+^2(\hbar\scale^2)}
 =
\frac{\pi^{1/4}}{\sqrt 2} (\hbar\scale^2)^{-3/4}
 \,,~~~~~~
N_\even = N_\odd = \frac{1}{\sqrt 2} N\,.
$$
In particular, $\Hi_+$ and $\Hi_-$ appear to be orthogonal.
Moreover, \eqref{GscaprohoHerm} yields $P_{-,n} = 0$ and $P_{0,n}
= P_{+,n}$, so that it suffices to determine $P_{+,n}$. Inserting
$\Sigma_+^r$ from \eqref{GSigma+r} into \eqref{GscaprohoHerm} and writing
$$
\ratio
 =
\sqrt\hbar\scale\tinco^{1/4} \equiv \sqrt{\frac{\scale}{g}}
$$
we obtain the identity
$$
\langle \psi_+,\xi_n^{(\infty)}\rangle
 =
(-1)^n \frac{\sqrt{(2n+1)!}}{2^n}
 \sum_{r=0}^n
 \frac{(-1)^r 2^{(2r+3)/2}}{r! (n-r)!}
\frac{\ratio^{3/2}}{(\ratio^2 + 1)^{(2n+3)/2}}\,.
$$
Taking the sum yields
 \beq\label{Gscaproprat}
\langle \psi_+,\xi_n^{(\infty)}\rangle
 =
(-1)^n \frac{\sqrt{(2n+1)!}}{2^n \, n!}
 \left(
\frac{2\ratio}{\ratio^2 + 1}
 \right)^{3/2}
 \left(
\frac{\ratio^2 - 1}{\ratio^2 + 1}
 \right)^n
\,.
 \eeq
Hence, in the harmonic oscillator approximation and the
approximation of $\theta_3$ by \eqref{Gapproxtheta}, the
expectation values $P_{+,n}$ are given by the rational functions
 \beq \label{GPprat}
P_{+,n}^{(\infty)}(\tau)
 =
\frac{(2n+1)!}{4^n \, (n!)^2}
 \left(
\frac{2\ratio}{\ratio^2 + 1}
 \right)^3
 \left(
\frac{\ratio^2 - 1}{\ratio^2 + 1}
 \right)^{2n}\,.
 \eeq
It is interesting to note that, in this approximation, $P_{+,n}$ depends on the
parameters $\hbar$, $\scale$ and $\nu$ only through the
ratio $\scale/\coco$. Figure \rref{AbbPprat} shows plots of $P_{+,n}^{(\infty)}$
and $P_{+,n}$ as functions of $\tinco$ on a logarithmic scale for
$\hbar\scale^2 = 1, \frac 1 2, \frac 1 8$ and $n=0,\dots,3$. We
see that for sufficiently small values of $\hbar\scale^2$ and
sufficiently small $n$ the approximation of $P_{+,n}$ by
$P_{+,n}^{(\infty)}$ is already satisfactory in the region of the dominant
maximum of $P_{+,0}$ and even more so for larger $\tau$. Hence, this
approximation can be used to study the behaviour of $P_{+,n}$ in this region.
Moreover, we claim that this approximation is consistent in the
sense that, for any $\ratio>0$,
$$
P_{+,n}^{(\infty)}(\ratio) \geq 0
 \,,~~~~~~
\sum_{n=0}^\infty P_{+,n}^{(\infty)}(\ratio) = 1
 \,.
$$
Indeed,
 \beq\label{GsumPp}
\sum_{n=0}^\infty P_{+,n}^{(\infty)}(\ratio)
 =
 \left(
\frac{2\ratio}{\ratio^2 + 1}
 \right)^3
\sum_{n=0}^\infty \frac{(2n+1)!}{4^n \, (n!)^2}
 \left(
\frac{\ratio^2 - 1}{\ratio^2 + 1}
 \right)^{2n}\,.
 \eeq
Recall that the function $y\mapsto (1-y)^{-3/2}$ has the Taylor series
$$
(1-y)^{-3/2} = \sum_{n=0}^\infty  \frac{(2n+1)!}{4^n \, (n!)^2}
~y^n\,,
$$
and this series is absolutely convergent for $|y|<1$. Replacing $y$ with
$(\ratio^2 - 1)^2/(\ratio^2 + 1)^2$, where $\tau>0$, we deduce that the
approximation is consistent in the asserted sense.

We determine the extremal points of $P_{+,n}^{(\infty)}$ on the positive
semiaxis. For $n=0$,
$$
\frac{\mr d}{\mr d \ratio}\,P_{+,0}^{(\infty)}(\ratio)
 =
\frac{24 \ratio^2(1-\ratio^2)}{(\ratio^2+1)^4}\,.
$$
Hence, at $\ratio = 1$, $P_{+,0}^{(\infty)}$ has a maximum, the maximal value
being
$$
P_{+,0}^{(\infty)} (\ratio = 1) = 1\,.
$$
This means that, for coupling constant $g=\scale$, up to the
approximations we have made, the state $\psi_+$ spanning $\Hi_+$
coincides with the ground state. In particular, the state $\psi_+$
is then approximately stationary. As remarked in Subsection
\rref{SSexpect}, the coincidence holds only in the approximation
and is not exact though. A physical interpretation of this
phenomenon has still to be found. For $n\geq 1$,
$$
\frac{\mr d}{\mr d \ratio}\,P_{+,n}^{(\infty)}(\ratio)
 =
 - \frac{(2n+1)!}{2^{2n-3} \, (n!)^2}~
\frac{\ratio^2(\ratio^2 - 1)^{2n-1}}{(\ratio^2 + 1)^{2n+4}}
 \left(
3 \ratio^4 - (8n+6) \ratio^2 + 3
 \right)\,.
$$
Hence, $P_{+,n}^{(\infty)}$ has maxima at
 \beq\label{GPpratmax}
\ratio_\pm
 =
 \sqrt{
\frac{4n + 3 \pm 2 \sqrt{4 n^2 + 6 n}}{3}
 }
 \eeq
and a minimum at $\ratio = 1$. The first maximum, $\ratio_-$, lies
in a region where the approximation is reliable only for very
small values of $\hbar\scale^2$, see Figure \rref{AbbPprat}. For
increasing $n$,
$\ratio_-$ approaches $\ratio=1$ from below and $\ratio_+$ moves towards
larger
values of $\ratio$. The maximal values of $P_{+,n}^{(\infty)}$ are
$$
P_{+,n}^{(\infty)}(\ratio_\pm)
 ~~=~~
\frac{3^{3/2} (2n+1)!}{2^{2n-3} \, (n!)^2}
 ~
 \frac{
\left(4n + 3 \pm 2 \sqrt{4 n^2 + 6 n}\right)^{3/2} \left(4n \pm 2
\sqrt{4 n^2 + 6 n}\right)^{2n}
 }{
\left(4n + 6 \pm 2 \sqrt{4 n^2 + 6 n}\right)^{2n+3}
 }\,.
$$
These values are independent of the parameters $\hbar$, $\scale$ and $\nu$ and
decrease for increasing $n$.

In the minimum $\ratio = 1$, $P_{+,n}^{(\infty)}$ vanishes. This
is consistent with what we have found for
$P_{+,0}^{(\infty)}$. The order of contact of $P_{+,n}^{(\infty)}$
with the real axis is $2n$. This order of contact is reflected in a broadening
of the valley between the two maxima, see Figure \rref{AbbPprat}.


\section{Outlook}


There is still more to say about the case of $\SU(2)$. The expectation values
$P_{\pm,n}$ and $P_{0,n}$ in the region between the strong and weak coupling
approximations and the dynamics relative to the costratified structure remain to
be studied. The exploration of that dynamics will rely on a detailed
investigation of the probability flow into and out of the subspaces $\Hi_\pm,
\Hi_\sing$.  

The next step is the construction of the costratified Hilbert space and the
subsequent analysis of various physical quantities for $\SU(3)$. Here, the orbit
type stratification of the reduced phase space has a 4-dimensional stratum, a
2-dimensional stratum, and three isolated points. Thereafter the construction
remains to be extended to an arbitrary lattice.

Finally, the costratified Hilbert space structure exploited in this paper
implements the stratification of the reduced classical phase space
on the level of states but leaves open the question what the stratification
might signify for the quantum observables, a question to be clarified in the
future. Then the physical role of this stratification can, perhaps, be
studied in more realistic models like lattice QCD \cite{qcd1,qcd2,qcd3}.


\section*{Acknowledgement}


The authors would like to express their gratitude to
Szymon Charzy\'nski, Heinz-Dietrich
Doebner, Alexander Hertsch, Jerzy Kijowski and Konrad Schm\"udgen for 
the stimulus of conversation,
to Brian Hall  for pointing out the relationship between the
states spanning $\Hi_\pm$ and coherent states, 
to Christian Fleischhack for hints at Thiemann's work on 
the overlap of coherent states,
and to Jim Stasheff for a number of comments
which helped improve the exposition.
J.\ H.\ and M.\ S.\ acknowledge funding by the German Research
Council (DFG) under contract \ Le 758/22-1 and contract RU692/3,
respectively.


\end{document}